\documentclass{svjour2}
\usepackage{graphicx}
\usepackage{subfigure}
\usepackage{ifthen}
\usepackage[mathscr]{euscript}

\newcount\myp

\usepackage{tikz}
\usetikzlibrary{decorations.pathmorphing}
\usetikzlibrary{decorations.markings}
\usetikzlibrary{arrows,shapes,snakes,automata,backgrounds,petri}
\usetikzlibrary{calc}
\usetikzlibrary{patterns}
\usetikzlibrary{decorations.text}

\tikzset{
xxtsubstrate/.style={decorate, 
line width=1pt,
draw=olive, 
decoration=snake, 
segment amplitude=0.75mm, 
line after snake=0.25mm,
line before snake=0.25mm
},
tsubstrate/.style={decorate, 
line width=1pt,
draw=olive, 
decoration=snake, 
segment amplitude=0.5mm, 
segment length=5pt,
segment amplitude=0.2mm, 
line after snake=1mm,
line before snake=1mm
},
Bsubstrate/.style={decorate, 
line width=1pt,
draw=olive, 
decoration=snake,
segment length=5pt,
segment aspect=0,
segment amplitude=0.5mm, 
line after snake=0mm,
line before snake=0mm
},
substrate/.style={decorate, 
line width=1pt,
draw=olive, 
decoration=snake, 
segment length=5pt,
segment amplitude=0.5mm, 
line after snake=0.5mm,
line before snake=0.5mm
},
activity/.style={very thick,draw=red,postaction={decorate},
decoration={markings,mark=at position .5 with
{\arrow[draw=red]{>}}}},
tactivity/.style={thick,draw=red,postaction={decorate},
decoration={markings,mark=at position .5 with
{\arrow[draw=red]{>}}}},
tEPSactivity/.style={thick,draw=red,postaction={decorate},
decoration={markings,mark=at position .55 with
{\arrow[draw=red]{>}}}},
tAactivity/.style={thick,draw=red},
Aactivity/.style={very thick,draw=red},
tSactivity/.style={thick,draw=red,postaction={decorate},
decoration={markings,mark=at position .7 with
{\arrow[draw=red]{>}}}},
Sactivity/.style={very thick,draw=red,postaction={decorate},
decoration={markings,mark=at position .7 with
{\arrow[draw=red]{>}}}}
}

\newcommand{\crossblob}[2][]{\coordinate (a) at (#2);
\draw[color=olive,line width=1pt,fill=white,text=black,outer sep=2pt] (a) circle (0.15cm) node [above] {#1};%
\draw[color=olive,line width=1pt] ($  (45:0.15cm) + (a) $) -- ($ (45:-0.15cm) + (a) $);%
\draw[color=olive,line width=1pt] ($ (-45:0.15cm) + (a) $) -- ($ (-45:-0.15cm) + (a) $);%
}

\usepackage{amsmath}
\usepackage{amsfonts}
\usepackage{cases}
\usepackage{xspace}
\usepackage{psfrag}
\usepackage{subfig}
\usepackage{color}

\newcommand{\ket}[1]{\left|#1\right\rangle}

\newcommand{\ave}[1]{\left\langle #1 \right\rangle}
\newcommand{\spave}[1]{\overline{#1}}

\newcommand{\imag}{\imath}

\newcommand{\plaind}{\mathrm{d}}
\newcommand{\dint}[1]{\mathchoice{\!\plaind#1\,}{\!\plaind#1\,}{\!\plaind#1\,}{\!\plaind#1\,}}
\newcommand{\ddint}[1]{\mathchoice{\!\plaind^d#1\,}{\!\plaind^d#1\,}{\!\plaind^d#1\,}{\!\plaind^d#1\,}}

\newcommand{\dbar}{\plaind\mkern-6mu\mathchar'26}
\newcommand{\deltabar}{\delta\mkern-6mu\mathchar'26}
\newcommand{\dintbar}[1]{\mathchoice{\!\dbar#1\,}{\!\dbar#1\,}{\!\dbar#1\,}{\!\dbar#1\,}}
\newcommand{\ddintbar}[1]{\mathchoice{\!\dbar^d#1\,}{\!\dbar^d#1\,}{\!\dbar^d#1\,}{\!\dbar^d#1\,}}
\newcommand{\ddMOneintbar}[1]{\mathchoice{\!\dbar^{d-1}#1\,}{\!\dbar^{d-1}#1\,}{\!\dbar^{d-1}#1\,}{\!\dbar^{d-1}#1\,}}

\usepackage{dsfont}

\newcommand{\gpset}[1]{\mathds{#1}}

\newcommand{\canetset}[1]{{\mathchoice {\hbox{$\sf\textstyle #1\kern-0.4em #1$}}
{\hbox{$\sf\textstyle #1\kern-0.4em #1$}}
{\hbox{$\sf\scriptstyle #1\kern-0.3em #1$}}
{\hbox{$\sf\scriptscriptstyle #1\kern-0.2em #1$}}}}

\newcommand{\Nset}{\gpset{N}}
\newcommand{\Zset}{\gpset{Z}}

\newcommand{\Rset}{\gpset{R}}

\def\nbZ{{\mathchoice {\hbox{$\sf\textstyle Z\kern-0.4em Z$}}
{\hbox{$\sf\textstyle Z\kern-0.4em Z$}}
{\hbox{$\sf\scriptstyle Z\kern-0.3em Z$}}
{\hbox{$\sf\scriptscriptstyle Z\kern-0.2em Z$}}}}

\newcommand{\gpvec}[1]{\mathbf{#1}}

\newcommand{\kvec}{\gpvec{k}}

\newcommand{\nvec}{\gpvec{n}}

\newcommand{\xvec}{\gpvec{x}}

\newcommand{\OC}{\mathcal{O}}
\newcommand{\PC}{\mathcal{P}}

\newcommand{\phitilde}{\tilde{\phi}}

\newcommand{\half}{\mathchoice{\frac{1}{2}}{(1/2)}{\frac{1}{2}}{(1/2)}}

\newcommand{\fourth}{\mathchoice{\frac{1}{4}}{(1/4)}{\frac{1}{4}}{(1/4)}}

\newcommand{\quarter}{\fourth}

\renewcommand{\exp}[1]{\mathchoice{e^{#1}}{\operatorname{exp}\left(#1\right)}{\operatorname{exp}\left(#1\right)}{\operatorname{exp}\left(#1\right)}}

\newcommand{\elabel}[1]{\label{eq:#1}}
\newcommand{\eref}[1]{(\ref{eq:#1})}
\newcommand{\Eref}[1]{Eq.~(\ref{eq:#1})}

\newcommand{\slabel}[1]{\label{sec:#1}}

\newcommand{\Sref}[1]{Section~\ref{sec:#1}}

\newcommand{\flabel}[1]{\label{fig:#1}}

\newcommand{\Fref}[1]{Figure~\ref{fig:#1}}

\newcommand{\latin}[1]{{\it #1}}
\newcommand{\ie}{\latin{i.e.}\@\xspace}
\newcommand{\eg}{\latin{e.g.}\@\xspace}
\newcommand{\cf}{\latin{cf.}\@\xspace}
\newcommand{\etc}{\latin{etc.}\@\xspace}

\newlength \standardfigwidth
\DeclareMathAlphabet{\matheub}{U}{eur}{m}{n}

\newenvironment{subeqnarray}[1]{\begin{subequations}#1\begin{eqnarray}}{\end{eqnarray}\end{subequations}\ignorespacesafterend}

\makeatletter
\newcommand{\creat}[3][]{\@ifempty{#1}{#2^{\dagger}}{\left(#2^{\dagger}\right)^{#1}}\@ifempty{#3}{}{\!(#3)}}

\newcommand{\creatDoi}[3][]{\@ifempty{#1}{\tilde{#2}}{\left(\tilde{#2}\right)^{#1}}\@ifempty{#3}{}{(#3)}}

\newcommand{\annih}[3][]{#2\@ifempty{#1}{}{^{#1}}\@ifempty{#3}{}{(#3)}}

\makeatother

\newcommand{\fnlabel}[1]{\label{fn:#1}}
\newcommand{\fnref}[1]{\ref{fn:#1}}

\newcommand{\MC}{\mathcal{M}}
\newcommand{\renorm}{\mathscr{R}}
\newcommand{\ai}[4]{{\left[\begin{smallmatrix}\textcolor{red}{#1}&\textcolor{red}{#2}\\\textcolor{olive}{#3}&\textcolor{olive}{#4}\end{smallmatrix}\right]}}

\newcommand{\aiXooo}[2]{\ifthenelse{#1=1}%
{\ai{#2}{\aiArr{2}}{\aiArr{3}}{\aiArr{4}}}%
{\ifthenelse{#1=2}%
{\ai{\aiArr{1}}{#2}{\aiArr{3}}{\aiArr{4}}}%
{\ifthenelse{#1=3}%
{\ai{\aiArr{1}}{\aiArr{2}}{#2}{\aiArr{4}}}%
{\ai{\aiArr{1}}{\aiArr{2}}{\aiArr{3}}{#2}}%
}}}
\newcommand{\aiX}[2]{\ifthenelse{#1=1}%
{\ai{#2}{\aiArrTWO}{\aiArrTHREE}{\aiArrFOUR}}%
{\ifthenelse{#1=2}%
{\ai{\aiArrONE}{#2}{\aiArrTHREE}{\aiArrFOUR}}%
{\ifthenelse{#1=3}%
{\ai{\aiArrONE}{\aiArrTWO}{#2}{\aiArrFOUR}}%
{\ai{\aiArrONE}{\aiArrTWO}{\aiArrTHREE}{#2}}%
}}}
\newcommand{\aiGammaX}[2]{\ifthenelse{#1=1}%
{\ai{#2}{\aiGammaArrTWO}{\aiGammaArrTHREE}{\aiGammaArrFOUR}}%
{\ifthenelse{#1=2}%
{\ai{\aiGammaArrONE}{#2}{\aiGammaArrTHREE}{\aiGammaArrFOUR}}%
{\ifthenelse{#1=3}%
{\ai{\aiGammaArrONE}{\aiGammaArrTWO}{#2}{\aiGammaArrFOUR}}%
{\ifthenelse{#1=4}%
{\ai{\aiGammaArrONE}{\aiGammaArrTWO}{\aiGammaArrTHREE}{#2}}%
{\ai{\aiGammaArrONE}{\aiGammaArrTWO}{\aiGammaArrTHREE}{\aiGammaArrFOUR}}%
}}}}
\newcommand{\aigammaX}[2]{\ifthenelse{#1=1}%
{\ai{#2}{\aigammaArrTWO}{\aigammaArrTHREE}{\aigammaArrFOUR}}%
{\ifthenelse{#1=2}%
{\ai{\aigammaArrONE}{#2}{\aigammaArrTHREE}{\aigammaArrFOUR}}%
{\ifthenelse{#1=3}%
{\ai{\aigammaArrONE}{\aigammaArrTWO}{#2}{\aigammaArrFOUR}}%
{\ifthenelse{#1=4}%
{\ai{\aigammaArrONE}{\aigammaArrTWO}{\aigammaArrTHREE}{#2}}%
{\ai{\aigammaArrONE}{\aigammaArrTWO}{\aigammaArrTHREE}{\aigammaArrFOUR}}%
}}}}
\newcommand{\aiArrONE}{-1}
\newcommand{\aiArrTWO}{-1}
\newcommand{\aiArrTHREE}{-1}
\newcommand{\aiArrFOUR}{-1}
\newcommand{\aiGammaArrONE}{-1}
\newcommand{\aiGammaArrTWO}{-1}
\newcommand{\aiGammaArrTHREE}{-1}
\newcommand{\aiGammaArrFOUR}{-1}
\newcommand{\aigammaArrONE}{-1}
\newcommand{\aigammaArrTWO}{-1}
\newcommand{\aigammaArrTHREE}{-1}
\newcommand{\aigammaArrFOUR}{-1}
\newcommand{\Gai}[4]{G^{\ai{#1}{#2}{#3}{#4}}}
\newcommand{\Gammaai}[4]{\Gamma^{\ai{#1}{#2}{#3}{#4}}}
\newcommand{\GammaaiR}[4]{\Gamma_{\renorm}^{\ai{#1}{#2}{#3}{#4}}}
\newcommand{\gammaai}[4]{\gamma^{\ai{#1}{#2}{#3}{#4}}}

\newcommand{\gammaTweak}{3}
\newcommand{\setgammaTweak}[1]{\renewcommand{\gammaTweak}{3}}

\newcommand{\Adim}{\mathtt{A}}
\newcommand{\Bdim}{\mathtt{B}}
\newcommand{\Ldim}{\mathtt{L}}
\newcommand{\Tdim}{\mathtt{T}}

\newcommand{\nBar}{\overline{n}}
\newcommand{\gammaBar}{\overline{\gamma}}
\newcommand{\kappaBar}{\overline{\kappa}}

\newcommand{\rhotilde}{\tilde{\rho}}
\newcommand{\dtilde}{\tilde{d}}
\newcommand{\Utilde}{\tilde{U}}
\newcommand{\epsilontilde}{\tilde{\epsilon}}
\newcommand{\psitilde}{\tilde{\psi}}
\newcommand{\Liouvillian}{\mathcal{L}}
\newcommand{\corresponding}{\hat{=}}
\newcommand{\epsnml}{\epsilon_{nm\ell}}

\begin{document}
\title{A field-theoretic approach to the Wiener Sausage}
\author{S Nekovar\thanks{On leave from University of Erlangen-N{\"u}rnberg} \and G Pruessner}
\institute{Department of Mathematics, 
Imperial College London, 
180 Queen's Gate,
London SW7 2AZ, 
United Kingdom\\
\email{g.pruessner@imperial.ac.uk}}
\maketitle

\begin{abstract}
The Wiener Sausage, the volume traced out by a sphere attached to a
Brownian particle, is a classical problem in statistics and mathematical
physics. 
Initially motivated by a range of field-theoretic, technical questions,
we present a single loop renormalised perturbation theory of a stochastic process closely related to the Wiener
Sausage, which, however, proves
to be exact for the exponents and
some amplitudes. 
The field-theoretic approach is particularly elegant and very enjoyable
to see at work on such a classic problem.
While we recover a number of known, classical results, the field-theoretic 
techniques deployed provide a particularly versatile framework,
which allows easy calculation with different boundary conditions even of
higher momenta and more complicated correlation functions. At the same
time, we provide a highly instructive, non-trivial example for some of
the technical particularities of the field-theoretic description of
stochastic processes, such as excluded volume, lack of translational
invariance and immobile particles.
The aim of the present work is not to improve upon the well-established
results for the Wiener Sausage, but to provide a field-theoretic
approach to it, in order to gain a better understanding of the 
field-theoretic obstacles to overcome.
\end{abstract}

\section{Introduction}
The Wiener Sausage problem \cite{KolmogoroffLeontowitsch:1933}
is concerned with determining the volume traced out (the sausage) by a $d$-dimensional
sphere attached to a Brownian particle in $d$ dimensions. The problem is
illustrated in \Fref{2d_example} in dimension $d=2$. It has been studied
extensively in the literature \cite{Montroll:1964,Spitzer:1964,KacLuttinger:1974,DonskerVaradhan:1975,LeGall:1986,BerezhkovskiiMakhnovskiiSuris:1989,vanWijlandCaserHilhorst:1997,Sznitman:1998,vandenBergBolthausendenHollander:2001,Spitzer:2001} from a probabilistic point of view and has
a very wide range of applications, such as 
medical physics \cite[for example]{DagdugBerezhkovskiiWeiss:2002}, 
chemical engineering \cite[for example]{EggersdorferPratsinis:2014}
or
ecology \cite[for example]{Visser:2007}.
On the lattice, the volume of the Sausage translates to the number of
\emph{distinct} sites visited \cite{Torney:1986}. In
this work, we present an alternative, field-theoretic approach which is particularly
flexible with respect to boundary conditions and observables 
with the aim to characterise and resolve the technical challenges in such an undertaking, not with the aim to improve upon the existing theory of the Wiener Sausage.

The approach has the additional appeal that, somewhat similar to percolation
\cite{StaufferAharony:1994} where all non-trivial features are due to the imposed definition
of clusters as being composed of occupied sites connected via open bonds
between nearest neighbours, the ``interaction'' in the present case is
one imposed
 in retrospect. After all,
the Brownian particle studied is free and not affected by any form of
interaction. Yet, the observable requires us to discount returns, \ie
loops in the trajectory of the particle, thereby inducing an interaction
between the particle's past and present.

\begin{figure}
\begin{center}
\includegraphics[width=0.5\linewidth]{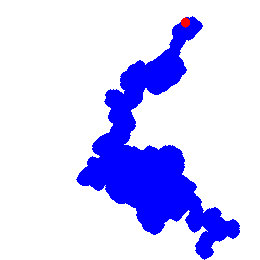}
\end{center}
\caption{\flabel{2d_example}Example of the Wiener Sausage problem in two
dimensions. The blue area has been traced out by the Brownian particle
attached to a disc
shown in red.}
\end{figure}

Before describing the process to be analysed in further detail, we want
to point out that some of the questions pursued in the following are
common to the field-theoretic re-formulation of stochastic processes
\cite{Doi:1976,Peliti:1985,Cardy:1999,Cardy:2008,TaeuberHowardVollmayr-Lee:2005,Taeuber:2014}. Against the
background of a field theory of the Manna Model
\cite{Manna:1991a,Dhar:1999c} one of us recently
developed, the features we wanted to understand were: 1) ``Fermionic'',
``excluded volume'' or ``hard-core interaction''
processes \cite[for example]{Hinrichsen:2000a}, \ie processes where lattice sites have a certain carrying
capacity (unity in the present case) that cannot be exceeded. 
2) Systems with boundaries, \ie lack
of momentum conservation in the vertices. 2') Related to that, how different
modes couple in finite, but translationally invariant systems (periodic boundary conditions). 3) The
special characteristics of the propagator of the immobile species.
4) Observables that are spatial or spatio-temporal integrals of
densities.

The Wiener Sausage incorporates all of the above and
because it is exactly solvable or has been characterised by very
different means
\cite{KolmogoroffLeontowitsch:1933,DonskerVaradhan:1975,BerezhkovskiiMakhnovskiiSuris:1989,vandenBergBolthausendenHollander:2004}, it also gives access to a better understanding of the
renormalisation process itself. 
In the following section we will describe the process we are
investigating and contrast it with the original Wiener Sausage. In
\Sref{FieldTheory} we will introduce the field-theoretic description up
to tree level, which is complemented by \Sref{beyond_tree_level}, where
we perform a one-loop renormalisation procedure. It will turn out that
there are no further corrections beyond one loop and our perturbative
results may thus be regarded as exhaustive. \Sref{semi_infty_strip} and
\Sref{infinite_cylinder} are dedicated to calculations in finite systems. \Sref{discussion} contains a discussion of the results mostly from a field-theoretic point of view, with \Sref{summary_results} however focusing on a summary of this work with regard to the original Wiener Sausage problem.

\section{Model}\slabel{model}
Originally, the Wiener Sausage is concerned with the moments or
generally statistical properties as a function of time of the
volume traced out by a sphere of fixed given (say, unit) radius, which is attached
to a Brownian particle. This volume is thus the set of point within a
certain distance to the particle's trajectory. Our field-theoretic
approach will not recover that process, but one that can reasonably be
assumed to reside in the same universality class. One may take the view
that the field-theoretic description is merely a different view on the
same phenomenon, namely the Wiener Sausage.

To motivate the field theory and link it to the original problem, we will 
distinguish three different models:
(i) The original Wiener Sausage in 
terms of a sphere dragged by a Brownian particle
\cite{KolmogoroffLeontowitsch:1933}, (ii) a discrete time random walker on a lattice, 
where the Sausage becomes the set of distinct sites visited
\cite{Torney:1986}, (iii) a Brownian particle 
in the continuum that spawns immobile offspring with a finite rate and
subject to a finite carrying capacity.
In the following, we will first describe how the phenomenon
on the lattice, (ii), relates to the original Wiener Sausage, (i), and then
how the field theory, (iii), relates to the lattice model, (ii).

The asymptote in long times $t$ of the number of distinct sites visited by 
a discrete time random walker on a lattice ((ii) above) is expected to converge to
that of the volume $V(t)$ over the volume $V_0$ of the sphere in the original
process ((i) above), provided the walker returns repeatedly, so that the shape of
the sphere and the structure of the lattice respectively do not enter
into the shape and size of the volume visited. 
Frequent returns are realised in the limit of long
times and below $d=2$ dimensions. 
In that case, the walker
on the lattice becomes a discretised version of the original Wiener Sausage,
as the particle drags a sphere that is small compared to the volume
traced out.
Indeed, in one dimension, $d=1$, the expected
volume of the Wiener Sausage in units of the volume of the sphere is
dominated by $V(t)/V_0 \sim \sqrt{4 tD/(\pi b^2)}$, where $t$ is the time, $D$ is the
diffusion constant and $b$ the radius of the sphere, whereas the
expected number of distinct sites visited by a random walker 
after $n$ steps is dominated $\sqrt{8n/\pi}$ \cite{Torney:1986}. The two
expressions are identical for $t=n$ and $D=2b^2$, the effective
diffusion constant of a random walker taking one step of distance $2b$
in one of the $2$ directions in each time step. Above $d=2$ dimensions, the walker is
free, \ie self-intersection of the trace becomes irrelevant on larger
time scales. The number of distinct sites visited and the Wiener Sausage
volume therefore both scale linearly in $t$ and $n$ respectively.
However, the (non-universal) proportionality factor, \eg
$\lim_{t\to\infty} V(t)/(V_0 t)$ for the original Wiener Sausage,
is affected by the microscopic details such as the self-intersection of
the sphere or the lattice structure of the random walker.

We proceed to relate the process on the lattice (ii) to a
Brownian particle spawning immobile offspring (iii). 
To this end, we first describe (ii) in the language of reaction and
diffusion. In (ii), an ``active'' particle (species ``A'', the active
species) performs a random walk on a lattice. Simultaneously, the
particle spawns immobile offspring particles (species ``B'', the blue
ink traces of A shown in \Fref{2d_example}, below sometimes
referred to as a ``substrate particle'') at every site visited, provided
that the site is not already occupied by an immobile B particle. In
other words, A spawns exactly one B at every newly visited site, so that
the number of B particles deposited becomes a proxy for the number of
distinct sites visited. In
dimensions less than $2$ the A particle will return to every site visited
arbitrarily often in the limit of long times. 
A finite spawning
probability
will therefore change the number of B particles deposited only at the
fringes of the set of sites visited, without, however, changing the
asymptotics of the number of B particles in the system as a function of
time. If $\nBar_B(\xvec,t)$ is the number of B particles at position $\xvec$
on the lattice and time $t$, the probability with which B particles are spawned by an A
particle may be written as $\gammaBar(1-\nBar_B(\xvec,t))$, so that deposition
occurs with probability $\gammaBar$ if no B particle is present and not at all
otherwise.

At this stage, we may introduce a \emph{carrying capacity} $\nBar_0$,
which determines the maximum number of B particles deposited on any
site, by making the spawning probability drop from $\gammaBar$
to $0$ linearly in the particle number $\nBar_B(\xvec,t)$, \ie like
\begin{equation}\elabel{def_carrying_capacity_lattice}
\gammaBar\frac{\nBar_0-\nBar_B(\xvec,t)}{\nBar_0} \ .
\end{equation}
In the process (ii) discussed so far, $\nBar_0$
is unity, but from what has been discussed above, $\nBar_0>1$ will result in
each (frequently) revisited site carrying $\nBar_0$ immobile B particles.
The meaning of the carrying capacity in relation to the field theory is
further discussed in \Sref{carrying_capcity}.

To see the relation between the third process, (iii) and the discrete
time, discrete space process (ii), we first introduce continuous time in
the latter. Random hopping which used to occur once in every time step
now becomes a Poisson process with a certain rate, say $H$, as does the
spawning of immobile offspring, with now takes place with rate
$\gamma(\nBar_0-\nBar_B(\xvec,t))/\nBar_0$. In the limit of $\gamma\gg H$ all
distinct sites visited will carry $\nBar_0$ immobile offspring. However, in
dimensions $d<2$ sites are visited repeatedly, so that even a finite
deposition (attempt) rate $\gamma$ yields the same asymptotic
occupation. In dimensions $d>2$ the number of B particles deposited
will, on the other hand, be proportional to the rate $\gamma$.

The expression $\gamma(\nBar_0-\nBar_B(\xvec,t))/\nBar_0$ may be written
as $\gamma - \kappaBar \nBar_B(\xvec,t)$ where $\kappaBar=\gamma/\nBar_0$ is a
discount rate. In this interpretation (the view adopted in the
perturbation theory below), deposition takes place unhindered
with rate $\gamma$ while \emph{unlimited} (and thus supposedly
suppressed) deposition is discounted by $\kappaBar \nBar_B(\xvec,t)$, \ie
with a rate proportional to the occupation and inversely proportional to
the carrying capacity.

It remains to take the continuum (space) limit to arrive at process
(iii) to be written as a field theory, where occupation numbers 
$\nBar_B(\xvec,t)$ and $\nBar_A(\xvec,t)$ for B and A particles
respectively turn into
occupation densities, \ie fields, namely 
$n_B(\xvec,t)$ and $n_A(\xvec,t)$.
Moreover, the carrying capacity $\nBar_0$ turns into
a carrying density capacity, $n_0$, so that the discount mentioned
above is now
parameterised by $\kappa=\gamma/n_0$ (a rate per density). The
deposition thus occurs with rate
\begin{equation}
\gamma\left(1-\frac{n_B(\xvec,t)}{n_0}\right) = \gamma - \kappa n_B(\xvec,t) \ .
\elabel{def_carrying_capacity}
\end{equation}
The random movement of the A particle is now parameterised by the
diffusion constant $D$, which may be obtained as the hopping rate $H$ over
the squared lattice spacing in the limit of the latter going to $0$.

\subsection{Intermediate summary}
The long-winded discussion above serves as a justification as to why we expect the
field theory of (iii) to produce a phenomenon in the same universality class as
the original Wiener Sausage. Starting from the original Wiener Sausage
(i), we have motivated why the process on the lattice, (ii), can be
regarded as a discretised version of (i) and introduced (iii) as its
continuum approximation.
In the course of the justification, we
made use of some of the details of the processes involved, such as
repeated returns to sites in process (ii). The field theoretic
description of the universality class of the Wiener Sausage, however,
may be derived without
recourse to these details, simply by observing that the volume traced
out by the sausage is proportional to the length of the trajectory with
multiple visits discounted, corresponding to the number of immobile
B particles deposited by a Brownian particle (of species A), if its spawning rate is
moderated down in the presence of B particles.

To summarise, process (iii), to be cast in a Liouvillian and thus a
field theory below,
is defined as follows: The Brownian particle A
freely diffuses with diffusion constant $D$ and possibly subject to
boundary conditions.  While diffusing, the particle
can spawn offspring with Poissonian rate $\gamma$ which, however, belong
to an immobile second species B. If $n_B(\xvec,t)$ is the density of
these particles, the deposition is linearly regulated down in their
presence, according to $\gamma-\kappa n_B(\xvec,t)$ with
$\kappa=\gamma/n_0$. Here, $\gamma$ is the deposition rate and $n_0$ is the
carrying (density) capacity.

It is convenient in the field theory to allow for spontaneous extinction 
with rate (or ``mass'') $r$.
Ignoring boundary conditions, the propagator of the Brownian particle
(species A, the ``activity'') takes the familiar form $1/(-\imag \omega + D \kvec^2
+ r)$ where $\omega$ and $\kvec$ parameterise frequency and momentum
(wave number) coordinates, respectively. The propagator of the immobile species takes the
form $1/(-\imag \omega + \epsilon')$, where $\epsilon'$ is the rate of
spontaneous extinction of $B$ therticles and the limit $\epsilon'\to0^+$
is implied to establish causality, as often done in field theories. 
The key observable, corresponding to the volume of the sausage, 
is the total number of immobile particles in the
system after a given time $t$, \ie the spatial integral over
$n_B(\xvec,t)$.

The engineering dimension of $\kappa$ is a rate per density, which in
comparison to the engineering dimension of the diffusion constant, an
area per time, reveals the upper critical dimension of $2$. 
Alternatively, this can be seen from the density of unhindered
deposition as a function of time, $(\gamma t)/(Dt)^{d/2}$, \ie in the
absence of discounts, $\kappa=0$ or $n_0\to\infty$.

In what follows, we will characterise process (iii) field theoretically.
The Liouvillian of the process is split into a linear part,
\Eref{harmonic_part}, discussed in \Sref{harmonic_part}, and
a non-linear part \Eref{nonlinear_part}, discussed in
\Sref{non-linearity}. The linear part of the Liouvillian can be
constructed from the propagators mentioned above and vice versa.
The Liouvillian will enter into a path-integral,
which can be used to generate all correlation and vertex functions. The
path integral itself is to be evaluated perturbatively in the
non-linearity, which, for example, instantly indicates that the
non-linearity has no bearing on the propagator of particle A. 

Above two dimensions, the non-linearity causes an ultraviolet divergence, \Eref{one_loop},
which has its origin
in the increasingly sharp divergence in $t$ of the density of a random
walker at the origin.\footnote{The integral \eref{one_loop} is in fact
the time-integrated density of a random walker at the origin, subject to
extinction $r+\epsilon'$, namely $\int_0^{\infty}\dint{t}
\exp{-(r+\epsilon')t}(4\pi Dt)^{-d/2}=\Gamma(1-d/2) (4\pi
D)^{-d/2}(r+\epsilon')^{-(1-d/2)}$.} However, in dimensions above $2$
the non-linearity is infrared irrelevant and so all long-time,
long-range observables are covered by the tree-level. In dimension below
$2$ no ultraviolet divergence occurs and the infrared can be regularised
using finite masses $r$ and $\epsilon'$. We will therefore work in
dimensions $2-\epsilon$, with $\epsilon>0$, known as dimensional
regularisation (of the ultraviolet).

Initially, the density fields will be studied on an infinite domain
without boundaries. However, in \Sref{semi_infty_strip} we will also
consider an infinite slab and in \Sref{infinite_cylinder} an infinite
cylinder. We will use Fourier transforms to write the fields in the
infinite domain and suitably chosen Fourier series 
in the presence of open (Dirichlet) or cylindrical boundaries. These
transforms and series are discussed in
\Sref{Fourier} and used later in the respective sections.

We will demonstrate in the following that the field theory recovers
exact results of the original Wiener Sausage as far as universal
exponents are concerned, but also with respect to some amplitudes
(namely the leading order term of the volume of the sausage in one
dimension as a function of time and the leading order of the volume as a
function of the system size of the infinite slab). Firstly, the present results
confirm that the logistic term \Eref{def_carrying_capacity} is capable
of capturing the constraints due to the carrying capacity. At a more
technical level, the
calculations for (partially) finite systems (infinite slab and cylinder)
involve different propagators, which under renormalisation can lead to new
non-linearities. Similar to the classic case discussed in
\cite{LubenskyRubin:1975} this problem, however, will be avoided.  The
results for these more complicated boundary conditions show very
interesting crossover behaviour. Finally, from a physical point of
view, it is particularly interesting that the infrared regularisation of
the immobile species, $\epsilon'$, a neccessary ingredient as to preserve causality in
the absence of diffusion, can in principle be used to regularise the
theory as a whole, \ie without the need of a particle mass $r$.

Before introducing the field theory of the present model in
\Sref{FieldTheory}, we discuss in the following
briefly the intricacies of the fermionic nature of the B particles.

\subsection{Finite carrying capacity}\slabel{carrying_capcity}
To fully understand the effect and consequences of the carrying
capacity, it is best to reconsider the process on the lattice. A
carrying capacity of $\nBar_0=1$ 
in \Eref{def_carrying_capacity_lattice}
switches off the deposition of B particles
in their presence in a rather dramatic fashion, implementing a constraint that is
normally referred to as \emph{fermionic}, because there is never more
than one B particle deposited on a site. Raising $\nBar_0$ allows the spawning
rate to drop linearly in the occupation in an otherwise \emph{bosonic} setup.
While this may raise suspicion and invite the criticism of a fudge,
as
demonstrated below, such a bosonic regularisation may be interpreted as
the fermionic case on a lattice with a particular connectivity, \ie the
attempted regularisation is the original, fermionic case in disguise,
suggesting that no such regularisation is needed.

Some authors \cite[and references therein]{vanWijland:2001} avoid terms like
\Eref{def_carrying_capacity_lattice} or \Eref{def_carrying_capacity} by
expanding a suitable expression for $\delta_{1,\nBar_B(\xvec,t)}$, a 
Kronecker $\delta$-function. \Eref{def_carrying_capacity_lattice} and
\Eref{def_carrying_capacity} are \emph{not} leading order
terms in an expansion. For $\nBar_0=1$ and before taking any other approximation (\eg 
continuous space and density or removing irrelevant terms in the field theory) a logistic term
like \eref{def_carrying_capacity_lattice} is a representation of the original process as exact as 
one involving the Kronecker $\delta$-function. For $\nBar_0>1$ a logistic term gives rise
to a model that may be strictly different compared to one with a sharp carrying capacity 
implemented by, say, a Heaviside step-function, $\theta(\nBar_0-\nBar_B(\xvec,t))$, but nonetheless
one that may be of equal interest.

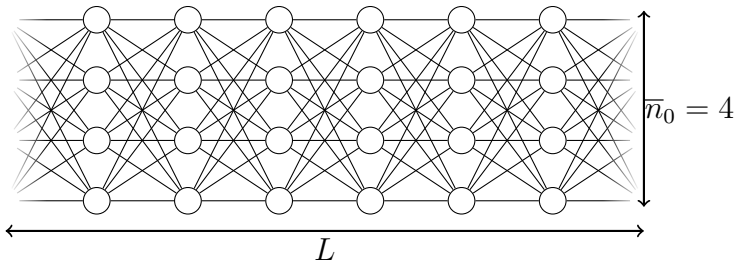
\begin{figure}
\begin{center}
\begin{tikzpicture}[scale=0.40]
\foreach \r in {1,3,5,7}
{
  \node[circle,fill=white,minimum size=10pt] (0-\r) at (0,\r) {};
  \node[circle,fill=white,minimum size=10pt] (21-\r) at (21,\r) {};
\foreach \c in {3,6,9,12,15,18}
{
  
  \node[draw,circle,minimum size=10pt] (\c-\r) at (\c,\r) {};
}
}
\foreach \t in {1,3,5,7}{
\foreach \r in {1,3,5,7}{
  \foreach \c in {0,3,6,9,12,15,18}{
  \myp=\c
  \advance\myp by 3
  \path (\c-\r) edge (\the\myp-\t);
  }
}
}
\pgfdeclarefading{faderight}
{\tikz\fill[left color=transparent!100,right
color=transparent!0] (0,0) rectangle (40bp,40bp); }
\pgfdeclarefading{fadeleft}
{\tikz\fill[left color=transparent!0,right
color=transparent!100] (0,0) rectangle (40bp,40bp); }
\fill[path fading=faderight,fill=white] (19.5,0) rectangle (21,8);
\fill[path fading=fadeleft,fill=white] (1.5,0) rectangle (0,8);
\draw[<->,thick](0,0) -- (21,0);
\node (L) at (10.5,-0.6) {\large $L$};
\draw[<->,thick](21,0.8) -- (21,7.3);
\node (n) at (22.5,4) {\large $\nBar_0=4$};
\end{tikzpicture}
\end{center}
\caption{\flabel{funny_lattice}A one dimensional lattice of size $L$ and carrying capacity
$\nBar_0=4$ corresponds to the lattice shown above, where the carrying capacity
of the former is implemented by expanding each site into a column of
$\nBar_0$
sites. The Brownian particle can jump from every site to all sites in the
neighbouring columns. In the new lattice, the carrying capacity \emph{per
site}
is unity, the carrying capacity \emph{per column} is $\nBar_0$.}
\end{figure}

Large $\nBar_0$ on the other hand, softens the cutoff, because 
spawning does not drop from suddenly from $\gammaBar$ to $0$ but is
more and more suppressed. One might
therefore be inclined to study the problem in the limit of large $\nBar_0$. 
At closer inspection, however, it turns out that such increased $\nBar_0$
does not present a qualitative change of the problem: Having $\nBar_0>1$ 
is \emph{as
if} each site was divided into $\nBar_0$ spaces. When the Brownian particle
jumps from site to site it arrives in one of those $\nBar_0$ spaces, only
$\nBar_0-\nBar_B(\nvec,t)$ of which are empty, so that an offspring can be left behind.
The process with carrying capacity $\nBar_0>1$ therefore corresponds to the
process with a carrying capacity of unity per space on a lattice where $\nBar_B(\nvec,t)$ describes the number
of immobile offspring in each ``nest'' or column of such spaces, as illustrated in
\Fref{funny_lattice}. In effect, the carrying capacity $\nBar_0>1$ is
implemented \emph{per column}, leaving the original fermionic constraint
of at most one offspring \emph{per space (or site)} in place.
In other words, even when a carrying
capacity $\nBar_0\gg1$ is introduced to smoothen the fermionic constraint, it is still
nothing else but the original constraint $\nBar_0=1$ on a different lattice.
This led us to believe that there is no qualitative difference in
$\nBar_0=1$ or any other finite value of $\nBar_0$. In the following,
the field theory will retain the carrying capacity
$n_0$ because it is an interesting parameter ($n_0\to\infty$ switches
the interaction off) and a ``marker'' of the
interaction. It may be set to any positive value.

\section{Field theory}
\slabel{FieldTheory}
In order to cast the model introduced above in a field-theoretic
language, we take the Doi-Peliti \cite{Doi:1976,Peliti:1985} approach without going
through too many technical details. There are a number
of reviews and extremely useful tutorials available
\cite{Cardy:1999,Cardy:2008}.

In the following the mobile particle is of species ``A'', performing
Brownian motion with (nearest neighbour) hopping rate $H$, which translates to diffusion constant 
$D=H/(2d)$ on a $d$-dimensional hypercubic lattice. We expect universal scaling in the large time and space limit.
To regularise the infrared,
we also introduce an extinction rate $r$. A's creation operator is
$\creat{a}{\xvec}$, its annihilation operator is $\annih{a}{\xvec}$.
The immobile species is ``B'', spawned with rate $\gamma$ by species A.
Its creation operator is
$\creat{b}{\xvec}$, its annihilation operator is $\annih{b}{\xvec}$,
both commuting with the creation and annihilation operators of species A.
The immobile species goes extinct with rate $\epsilon'$, which allows us
to have a Fourier transform and to restore causality (possible annihilation, \ie existence, only after
creation) even without spontaneous extinction, once we take the limit $\epsilon'\to0$. 

\subsection{Fourier Transform}\slabel{Fourier}
After replacing the operators by real fields, the Gaussian (harmonic)
part of the resulting path integral can be performed, once the fields
have been Fourier transformed. We will use the sign and notational
convention of 
\begin{multline}\elabel{fourier_transform}
\phi(\xvec,t) \\
= 
(2\pi)^{-1} \int_{-\infty}^{\infty} \dint{\omega}
(2\pi)^{-1} \int_{-\infty}^{\infty} \dint{k_1}\ldots
(2\pi)^{-1} \int_{-\infty}^{\infty} \dint{k_d}
\ \phi(\kvec,\omega) \exp{-\imag\omega t + \imag\kvec\xvec}\\
=
\int \dintbar{\omega}\ddintbar{k} 
\ \phi(\kvec,\omega) \exp{-\imag\omega t + \imag\kvec\xvec} \ .
\end{multline}
The field $\phi(\kvec,\omega)$ corresponds to the annihilator
$\annih{a}{\xvec}$ of the active particles, the field $\phitilde(\kvec,\omega)$ to the
Doi-shifted creator $\creatDoi{a}{\xvec}=\creat{a}{\xvec}-1$.
Correspondingly, $\psi(\kvec,\omega)$ and $\psitilde(\kvec,\omega)$
replace $\annih{b}{\xvec}$ and $\creatDoi{b}{\xvec}=\creat{b}{\xvec}-1$,
respectively. 

It is instructive to consider a second set of orthogonal
functions at this stage.
Placing the process in a space that has a \emph{finite extension} along
one axis means that
boundary conditions have to be met, which is more conveniently done in
one eigensystem rather than another. Below, we will consider
an infinite slab with finite thickness $L$, \ie
$d$-dimensional spaces which are infinitely
extended (using the orthogonal functions and transforms introduced
above) in $\dtilde=d-1$ dimensions, while along one axis, the boundaries are open, \ie the particle
density of species A vanishes at the (two parallel, $\dtilde$-dimensional) boundaries and outside. This Dirichlet boundary condition
is best met using eigenfunctions $\sqrt{2/L} \sin(q_n z)$ with $q_n=\pi
n/L$ 
and $n=1,2,\ldots$, making it complete and orthonormal because
\begin{equation}\elabel{efuncs_sin}
\frac{2}{L} \int_0^L\dint{z}\, \sin(q_n z) \sin(q_m z) = \delta_{n,m} \ .
\end{equation}
In passing, we have introduced
the finite linear length of the space, $L$. 
Purely for ease of notation and in order to keep expressions in finite
systems dimensionally as similar as possible to those in infinite ones,
\Eref{fourier_transform}, we will transform as follows:
\begin{subeqnarray}{\elabel{convenient_choices}}
\phi(z)&=&\frac{2}{L}\sum_{n=1}^\infty \sin(q_n z) \phi_n\\
\phi_n&=&\int_0^L \dint{z} \sin(q_n z) \phi(z)
\end{subeqnarray}
using
\begin{subeqnarray}{}
\frac{2}{L}\sum_{n=1}^\infty \sin(q_n y) \sin(q_n z) & = & \delta(z-y)\\
\int_0^L \dint{z} \sin(q_n z) \sin(q_m z) & = & \frac{L}{2} \delta_{m,n} \ ,
\elabel{two_terms_Dirichlet}
\end{subeqnarray}
where $\delta(z-y)$ is the usual Dirac $\delta$ function for
$z-y\in(0,L)$ but to be replaced by the periodic Dirac comb 
$\sum_{m=-\infty}^\infty \delta(z-y+mL)$ for arbitrary $z-y$. For ease
of notation, we have omitted the time dependence of $\phi(\xvec,t)$ as
well as $\dtilde$ components other than $z$. The other fields, $\phitilde$, 
as well as $\psi$ and $\psitilde$ transform correspondingly. The
spatial transform of the latter is subject to some convenient choice,
because the immobile species is not constrained by a boundary condition.

It will turn out that, as
expected in finite size scaling, the lowest mode $q_1=\pi/L$ plays the
r{\^o}le of a temperature like variable, controlling the distance to the
critical point. 

We will also briefly study systems which are infinitely extended in
$\dtilde$ dimensions and periodically closed in one.
In the periodic dimension, the spectrum of conveniently chosen
eigenfunctions $\sqrt{1/L} \exp{\imag k_n y}$ is discrete with
$k_n=2\pi n/L$ and $n\in\Zset$,
\begin{equation}\elabel{efuncs_exp}
\frac{1}{L} \int_0^L\dint{y}\, \exp{\imag k_n y} \exp{\imag k_m y} =  \delta_{n+m,0} \ .
\end{equation}
Again, we transform slightly asymmetrically (in $L$),
\begin{subeqnarray}{\elabel{convenient_choices_periodic}}
\phi(z)&=&\frac{1}{L}\sum_{n=-\infty}^\infty \exp{\imag k_n z} \phi_n\\
\phi_n&=&\int_0^L \dint{z} \exp{-\imag k_n z} \phi(z)
\end{subeqnarray}
with
\begin{subeqnarray}{}
\frac{1}{L}\sum_{n=-\infty}^\infty \exp{\imag k_n y} \exp{-\imag k_n z} & = & \delta(z-y)
\elabel{exp_sum}\\
\int_0^L \dint{z} \exp{\imag k_n z} \exp{-\imag k_m z} & = & L \delta_{m,n} \ ,
\end{subeqnarray}
where again $\delta(z-y)$ is to be replaced by a Dirac comb if considered for $z-y\notin
(0,L)$. Again, time and $\dtilde$ spatial coordinates were omitted.
Similar transforms apply to the other fields.

There is a crucial difference between eigenfunctions $\exp{\imag k_ny}$ and
$\sin(q_nz)$, as the former conserves momenta in vertices, whereas the
latter does not: 
\begin{subequations}
\elabel{delta_versus_epsilon_nml_all}
\begin{equation}
\int_0^L \dint{y}
\exp{\imag k_n y} \exp{\imag k_m y} \exp{\imag k_{\ell} y}  =  
L \delta_{n+m+\ell,0} \elabel{three_terms_periodic} 
\end{equation}
while
\begin{equation}
\int_0^L \dint{y} 
\sin(q_n y) \sin(q_m y) \sin(q_{\ell} y) = L \epsnml \ ,
\elabel{three_terms_Dirichlet}
\end{equation}
where
\begin{equation}
\epsnml \elabel{def_epsilon_nml}
=
\left\{
\begin{array}{ll}
\frac{1}{2\pi} \left(
\frac{1}{n+m-\ell}
+\frac{1}{n-m+\ell}
+\frac{1}{-n+m+\ell}
-\frac{1}{n+m+\ell}
\right)
& \text{for $\ell+m+n$ odd}\\
\quad & \quad \\
0            & \text{otherwise} 
\end{array}
\right.
\end{equation}
\end{subequations}
with $q_n=\pi n/L>0$, $n\in\Nset^+$ and $k_n=2\pi n/L$, $n\in\Zset$ (sign unconstrained) as
introduced above.

Having made convenient choices such as \Eref{convenient_choices}, we
will carry on using the Fourier transforms of the bulk
\Eref{fourier_transform}, which is
easily re-written for Dirichlet boundary conditions using
\Eref{convenient_choices},
simply by replacing each integral over $\dintbar{k}$ by $(2/L)\sum_n$
and similar for periodic boundary conditions,
\Eref{convenient_choices_periodic}. Only the non-linearity,
\Sref{non-linearity}, is expected to require further careful analysis as
$\epsnml$ of \Eref{three_terms_Dirichlet} is structurally far more demanding than
$\delta_{n+m+\ell,0}$ of \Eref{three_terms_periodic}.

\subsection{Harmonic Part}\slabel{harmonic_part}
Following the normal procedure \cite[for example]{Cardy:1998}, the harmonic
part $\Liouvillian_0$ of the Liouvillian $\Liouvillian=\Liouvillian_0+\Liouvillian_1$ reads
\begin{equation}\elabel{harmonic_part}
\Liouvillian_0 = 
\phitilde \partial_t \phi 
+ D \nabla \phitilde \nabla \phi 
+ r \phitilde \phi
+ \psitilde \partial_t \psi
+ \epsilon' \psitilde \psi \ .
\end{equation}
The non-linear part $\Liouvillian_1$, \Eref{nonlinear_part}, is
discussed in \Sref{non-linearity}. The harmonic part, $\Liouvillian_0$,
describes the diffusive evolution of the density field of A particles,
represented by $\phi$ and $\phitilde$,
which diffuse with diffusion constant $D$ and get
spontaneously extinct with rate $r$, as well as the evolution of immobile
particles B, represented by densities $\psi$ and $\psitilde$, which do
not diffuse but get extinct with rate $\epsilon'$.

After Fourier transforming and without further ado the harmonic part of the path integral 
\[
\int
\mathcal{D}\phi
\mathcal{D}\phitilde
\mathcal{D}\psi
\mathcal{D}\psitilde
\exp{-\int \ddintbar{\!k}\,\dintbar{\omega}\ \Liouvillian}
\]
can be performed, producing the two bare propagators
\begin{subequations}\elabel{bare_propagator}
\begin{align}\elabel{bare_activity_propagator}
\ave{\phi(\kvec,\omega)\phitilde(\kvec',\omega')}_0
&=
\frac{\deltabar(\kvec+\kvec')\deltabar(\omega+\omega')}{-\imag\omega+D\kvec^2+r}
\ \corresponding\ \tikz[baseline=-2.5pt]{\draw[activity] (0.6,0) -- (-0.6,0);}\\
\ave{\psi(\kvec,\omega)\psitilde(\kvec',\omega')}_0
&=
\frac{\deltabar(\kvec+\kvec')\deltabar(\omega+\omega')}{-\imag\omega+\epsilon'}
\ \corresponding\ \tikz[baseline=-2.5pt]{\draw[substrate] (0.6,0) -- (-0.6,0);}
\elabel{substrate_propagator}
\end{align}
\end{subequations}
where $\deltabar(\omega+\omega')=\delta(\omega+\omega')/(2\pi)$ and 
$\deltabar(\kvec+\kvec')=\delta(\kvec+\kvec')/(2\pi)^d$. 
Below, we will refer to the propagator
of the diffusive particles as the ``activity propagator'' and to the one
for the immobile species as the ``substrate propagator'' (or
``activity'' and ``substrate legs'', respectively).
As the propagation of the active particles is unaffected by the
deposition of immobile particles, the activity propagator does not
renormalise
$\ave{\phi\phitilde}=\ave{\phi\phitilde}_0$. The same is true for the
immobile species, which might be spawned by active particles, however,
once deposited remains inert, $\ave{\psi\psitilde}=\ave{\psi\psitilde}_0$.

The Fourier
transform \Eref{fourier_transform} of the latter produces
$\delta(\xvec-\xvec')\theta(t-t')$ in the limit $\epsilon'\to0$, with $\theta(x)$ denoting the
Heaviside $\theta$-function as one would expect (with $\xvec,t$
being the position and time of ``probing'' and $\xvec',t'$ position and time of 
creation).
At this stage, there is no interaction and no transmutation,
$\ave{\psitilde(\kvec,\omega)\phi(\kvec',\omega')}=0$. Diffusing
particles A happily co-exist with immobile ones.

\subsection{Non-Linearity}\slabel{non-linearity}
The harmonic part of the Liouvillian, $\Liouvillian_0$, discussed in the preceding section
covers the diffusive motion and spontaneous extinction of A particles (fields $\phi$ and
$\phitilde$) and the spontaneous extinction of the resting B particles
(fields $\psi$ and
$\psitilde$). In the following, we will discuss the non-linear
(interacting) part of the Liouvillian, $\Liouvillian_1$, which
introduces the spawning of B particles by the A particle, subject to the
constraint of the finite carrying capacity, which establishes an effective
interaction between
previously deposited
particles and any new particle to be deposited.

As discussed in \Sref{carrying_capcity}, spawning is moderated down in the
presence of B particles to $\gamma(1-n_B(\xvec,t)/n_0)$. At the level of
a master equation, this conditional deposition gives a non-linear contribution of
\begin{multline}\elabel{master_eqn}
\partial_t \PC(\ldots,n_A,n_B,\ldots) = \text{harmonic terms} +\\
  \sum_{\nvec} n_A \gamma\left(1-\frac{n_B-1}{n_0}\right)
  \PC(\ldots,n_A,n_B-1,\ldots)\\
- n_A \gamma\left(1-\frac{n_B}{n_0}\right) \PC(\ldots,n_A,n_B,\ldots) \ ,
\end{multline}
where, for convenience, the problem is considered for individual lattice
sites $\nvec$ which contain $n_A=n_A(\nvec)$ particles of species A and
$n_B$ particles of species B. The contributions by harmonic terms,
namely diffusion of A particles and spontaneous extinction of both, as
discussed in the previous section, have been omitted. The first term in the sum describes the
creation of a B particle in the presence of $n_B-1$ of those to make up $n_B$ in total, the second term makes the
B particle number exceed $n_B$, $n_B\to n_B+1$. If 
\begin{equation}
\ket{\Psi}(t)=\sum_{\{n_A,n_B\}} 
\PC(\ldots,n_A,n_B,\ldots) 
\prod_{\nvec} \creat[n_A]{a}{\nvec}
\prod_{\nvec} \creat[n_B]{b}{\nvec}
\ket{0} \ ,
\end{equation}
where the sum runs over all states of the entire lattice, then the
conditional deposition produces the contribution
\begin{multline}\elabel{non-linearities}
\partial_t \ket{\Psi}(t)=\text{bilinear terms} + \\
\sum_{\nvec}
 \gamma\ \creatDoi{b}{\nvec}\,\creat{a}{\nvec}\annih{a}{\nvec}
-\frac{\gamma}{n_0}\ \creatDoi{b}{\nvec}\creat{b}{\nvec}\annih{b}{\nvec}\,\creat{a}{\nvec}\annih{a}{\nvec}
\ ,
\end{multline}
where we have used the commutator,
$(\creat{b}{}\annih{b}{}-1)\creat{b}{}=\creat[2]{b}{}\annih{b}{}$ and 
the Doi-shifted creation operator, $\creat{b}{}=\creatDoi{b}{}+1$, as
well as the particle number operator $\creat{b}{}\annih{b}{}$.

Although using Doi-shifted operators throughout gives rise to a rather
confusing six non-linear
vertices, the resulting field theory does not turn out as messy as one
may expect. However, we need to allow for different renormalisation,
therefore introducing six different couplings below.

Replacing $\creat{a}{}$ by $1+\creatDoi{a}{}$ in
the first term of the sum generates the bilinearty
$\annih{a}{}\creatDoi{b}{}$, which we will parameterise in the following
by $\tau$, corresponding to a transmutation of
an active particle to an immobile one. Transmutation is obviously spurious; it does not actually take place but will allow us in the Doi-shifted setup (and thus with the corresponding left vacuum \cite{Cardy:1999,Cardy:2008}) to probe for substrate particles (using $\annih{b}{}$) after creating an active one (using $\creat{a}{}$) without having to probe for the latter (using $\annih{a}{}$). There is no advantage in moving
that to the bilinear part $\Liouvillian_0$, because the determinant of
the bilinear matrix $M$ in
\begin{equation}
\Liouvillian'_0=
\left(
\begin{array}{l}
\phitilde\\
\psitilde
\end{array}
\right)^T
\underbrace{
\left(
\begin{array}{cc}
-\imag \omega + D \kvec^2 + r & 0 \\
-\tau & -\imag \omega + \epsilon' 
\end{array}
\right)}_{M}
\left(
\begin{array}{l}
\phi\\
\psi
\end{array}
\right)
\end{equation}
is unaffected by $\tau\ne0$ and therefore none of the propagators
mentioned above change. One may therefore treat \emph{all} terms
(including the bilinear transmutation) resulting
from the interaction
perturbatively, with transmutation
\begin{multline}\elabel{deri_transmutation}
\ave{\psi(\kvec,\omega)\phitilde(\kvec',\omega')}_0\\
=\int \ddintbar{\!k''}\, \dintbar{\omega''} 
\ \ave{\psi(\kvec,\omega)\psitilde(\kvec'',\omega'')}
\tau
\ave{\phi(\kvec'',\omega'')\phitilde(\kvec',\omega')}\\
=\frac{\deltabar(\kvec+\kvec')\deltabar(\omega+\omega')}{-\imag\omega+\epsilon'}
\ \tau \ \frac{1}{-\imag\omega+D\kvec^2+r}
\ \corresponding\ \tikz[baseline=-2.5pt]{
\draw[activity] (0.6,0) -- (0,0) node[at end,above] {$\tau$};
\draw[substrate] (0,0) -- (-0.6,0);
}
\end{multline}
that is present regardless of the carrying capacity $n_0$. At this stage
it is worth noting the sign of $\tau$ (and $\sigma$ below) as positive,
\ie the perturbative expansion will generate terms with pre-factors
$\tau$ (and $\sigma$ below).

The only 
other non-linearity independent from the carrying capacity $n_0$
is the vertex
$\creatDoi{b}{}\creatDoi{a}{}\annih{a}{}$ (or $\psitilde\phitilde\phi$) in the following parameterised
by the coupling constant $\sigma$. Diagrammatically, it may be written
as the (amputated vertex)
\begin{equation}\elabel{sigma_diagram}
\tikz[baseline=-2.5pt]{
\draw[Aactivity] (0,0) -- (-0.4,0) node [at start,above] {$\sigma$};
\draw[Aactivity] (0.4,0) -- (0,0);
\draw[substrate] (-130:0.4) -- (0,0);
} \ ,
\end{equation}
and can be thought of as spawning, rather than transmutation parameterised
by $\tau$.

According to \Eref{non-linearities}, there are 
four  non-linearities with bare-level couplings of $\gamma/n_0$, 
generated by replacing the regular creation operators by their
Doi-shifted counterparts, 
$\creat{a}{\nvec}=1+\creatDoi{a}{\nvec}$ and 
$\creat{b}{\nvec}=1+\creatDoi{b}{\nvec}$,
in
$\frac{\gamma}{n_0}\ \creatDoi{b}{\nvec}\creat{b}{\nvec}\annih{b}{\nvec}\,\creat{a}{\nvec}\annih{a}{\nvec}$.
Each spawns at least one substrate particle, but more importantly, it
also annihilates at least one substrate particle as it ``probes for'' its
presence. The two simplest and most important (amputated) vertices are the ones introduced above with a
``wriggly tail added'',
\begin{equation}\elabel{mu_kappa}
\tikz[baseline=-2.5pt]{
\draw[Aactivity] (0.5,0) -- (0,0) node[at end,above] {$\lambda$};
\draw[substrate] (0,0) -- (-0.5,0);
\draw[substrate] (-50:0.5) -- (0,0);
}
\qquad
\tikz[baseline=-2.5pt]{
\draw[Aactivity] (0.5,0) -- (0,0) node[at end,above] {$\kappa$};
\draw[Aactivity] (0,0) -- (-0.5,0) node[at end,above] {};
\draw[substrate] (-130:0.5) -- (0,0);
\draw[substrate] (-50:0.5) -- (0,0);
}
\end{equation}
where we have also indicated their coupling. By mere inspection, it is
clear that those two vertices can be strung together, renormalising the
left one. In fact, $\kappa$ is the one and only coupling that
renormalises all non-linearities ($\sigma$,$\lambda$,$\kappa$,$\chi$ and
$\xi$), including itself.

Two more vertices are generated,
\begin{equation}\elabel{rho_nu}
\tikz[baseline=-2.5pt]{
\draw[Aactivity] (0.5,0) -- (0,0) node[at end,above] {$\ \ \chi$};
\draw[substrate] (0,0) -- (-0.5,0);
\draw[substrate] (-50:0.5) -- (0,0);
\draw[substrate] (130:0.5) -- (0,0);
}
\qquad
\tikz[baseline=-2.5pt]{
\draw[Aactivity] (0.5,0) -- (0,0) node[at end,above] {$\ \ \xi$};
\draw[Aactivity] (0,0) -- (-0.5,0) node[at end,above] {};
\draw[substrate] (-130:0.5) -- (0,0);
\draw[substrate] (-50:0.5) -- (0,0);
\draw[substrate] (130:0.5) -- (0,0);
} \ ,
\end{equation}
which become important only for higher order correlation functions of
the substrate particles, because there is no vertex annihilating more
than one of them --- correlations between substrate particles are present but not relevant
for the dynamics. Notably, there is no vertex that has more incoming
than outgoing 
substrate legs. Finally, we note that the sign with which $\lambda$,
$\kappa$, $\chi$ and $\xi$ are generated in the perturbative expansion
is negative.

For completeness, we state the interaction part of the Liouvillian (see
\Eref{harmonic_part})
\begin{equation}\elabel{nonlinear_part}
\Liouvillian_1 = 
- \tau \psitilde \phi 
- \sigma \psitilde \phitilde \phi
+ \lambda \psitilde \psi \phi
+ \kappa \phitilde \psitilde \psi \phi
+ \chi \psitilde^2 \psi \phi
+ \xi \phitilde \psitilde^2 \psi \phi \ ,
\end{equation}
with 
\begin{equation}
\tau=\sigma=\gamma 
\qquad \text{and} \qquad
\lambda=\kappa=\chi=\xi=\gamma/n_0
\elabel{bare_level_couplings}
\end{equation}
at bare level.

\newcommand{\dimensionof}[1]{\left[#1\right]}

\subsection{Dimensional analysis}
Determining the engineering dimensions of the coupling introduced above
is part of the ``usual drill'' and will allow us to determine the upper
critical dimension and to remove irrelevant couplings. Without dwelling
on details, analysis of the harmonic part, \Eref{harmonic_part}, 
reveals that $\dimensionof{D}=\Ldim^2/\Tdim$ (as expected for a diffusion
constant) and
$\dimensionof{r}=\dimensionof{\epsilon'}=1/\Tdim$ (as expected for all extinction rates), with
$\dimensionof{x}=\Ldim$, a length,
and $\dimensionof{t}=\Tdim$, a time. In real time and real space,
$\dimensionof{\phitilde\phi}=\dimensionof{\psitilde\psi}=\Ldim^{-d}$.

Performing the Doi-shift in \Eref{non-linearities} first and introducing
couplings for the non-linearities as outlined above allows for two
further independent dimensions, say spawning
$\dimensionof{\sigma}=\Adim$ and transmutation $\dimensionof{\tau}=\Bdim$ (both originally
equal to the rate $\gamma$), 
which implies 
$\dimensionof{\lambda}=\Adim^{-1}\Bdim \Ldim^d\Tdim^{-1}$, 
$\dimensionof{\kappa}=\Ldim^d\Tdim^{-1}$, 
$\dimensionof{\chi}=\Ldim^d\Bdim$, 
$\dimensionof{\xi}=\Ldim^d\Adim$, 
as well as
$\dimensionof{\psi}=\Tdim\Adim\Ldim^{-d}$,          $\dimensionof{\psitilde}=\Adim^{-1}\Tdim^{-1}$,
$\dimensionof{\phi}=\Adim \Bdim^{-1}\Ldim^{-d}$, $\dimensionof{\phitilde}=\Adim^{-1}\Bdim$ in
real space and time.

As far as the field theory is concerned, the only constraint is to
retain the diffusion constant on large scales, which implies $\Tdim=\Ldim^2$.
As a result, the non-linear coupling $\kappa$ (originally
$\gamma/n_0$) becomes irrelevant in dimensions $d>d_c$, as
expected with upper critical dimension $d_c=2$. 
The two independent engineering dimensions $\Adim$ and $\Bdim$ will be used in
the analysis below in order to maintain the existence of the associated
processes of transmutation and spawning, which are expected to govern
the tree level. If we were to argue that they become irrelevant above
a certain upper critical dimension, the density of offspring and its
correlations would
necessarily vanish everywhere.\footnote{Strictly, as we will demonstrate
below, $n$-point correlation
functions can be constructed with $\tau$ only, say
$\tikz[baseline=-2.5pt]{
\draw[tSactivity] (0.6,0) -- (0,0) node[at end,above] {$\tau$};
\draw[tsubstrate] (0,0) -- (-0.6,0);
\draw[tsubstrate] (-0.8,0) -- (-1.4,0);
\crossblob{-0.7,0};
}$
in
\Eref{second_order_diagrams}. However, it is clear that the density of
the active walker and its immobile offspring will remain correlated,
which is mediated by $\sigma$, \Eref{sigma_diagram}.}

Even though we may want to exploit the ambiguity in the engineering dimensions
\cite{LeeCardy:1995,TaeuberHowardVollmayr-Lee:2005} in the scaling
analysis (however, consistent with the results above),
in the following section we will make explicit use of the
Doi-shift when deriving observables, which means that both $\phitilde$ and $\psitilde$ are
dimensionless (in real space and time),
$\dimensionof{\phitilde}=\dimensionof{\psitilde}=1$, which implies
$\Adim=\Tdim^{-1}$ and $\Adim=\Bdim$. 
As expected, $\tau$ is then a rate (namely the
transmutation rate) and so is $\sigma$, 
$\dimensionof{\tau}=\dimensionof{\sigma}=1/\Tdim$. Also not unexpectedly, the
remaining four couplings all end up having the same engineering
dimension, 
$\dimensionof{\lambda}=\dimensionof{\kappa}=\dimensionof{\chi}=\dimensionof{\xi}=\Ldim^d\Tdim^{-1}$,
as suggested by $\gamma/n_0$, which is a rate
per density ($\gamma$ being the spawning rate and $n_0$ turning into a
carrying capacity density as we take the continuum limit).

\subsection{Observables at tree level: Bulk}\slabel{tree_level}
The aim of the present work is to characterise the volume of the Wiener
Sausage field-theoretically. 
As discussed in \Sref{model}, this is done not in terms of an actual spatial
volume, but rather in terms of the \emph{number} of spawned immobile
offspring.
In this section, we define the relevant observables in terms
of the fields introduced above. This is best done at tree level,
presented in the following, before
considering loops and the subsequent renormalisation. While the
tree level is the theory valid above the upper critical dimension, it is
equivalently the theory valid in the absence of any physical
interaction, \ie the theory of $n_0\to\infty$. We introduce the
observables first in the presence of a mass $r$, which amounts to
removing the particle after a time of $1/r$ on average.

If $v^{(1)}(\xvec;\xvec^*)$ is the density of substrate particles at $\xvec$ in a
particular realisation of the process at the end of
the life time of the diffusive particle which started at $\xvec^*$, the volume of the Sausage is
$V=\int\ddint{x} v^{(1)}(\xvec;\xvec^*)$. The ensemble average is then just 
$\ave{V}=\int\ddint{x} \ave{v^{(1)}(\xvec;\xvec^*)}$, where
$\ave{\bullet}$ 
denotes the ensemble average of $\bullet$ and the dependence on
$\xvec^*$ drops out in the bulk. Alternatively (as done below), one may
consider a distribution\footnote{This is a distribution of
experiments with \emph{one} initial particle each, rather than a
``smeared out'' initial particle, whose ``constituents'' would be able
to ``see'' each other's traces.} $d(\xvec^*)$ of initial starting points
$\xvec^*$, over which 
an additional expectation, denoted by an overline, $\spave{\bullet}$, has to be taken.

Higher moments require higher order correlation functions
\begin{equation}
\ave{V^n} = \int\ddint{x_1}\ldots\ddint{x_n} 
\ave{\spave{v}^{(n)}(\xvec_1,\ldots,\xvec_n)} \ ,
\end{equation}
where 
\begin{equation}
\ave{\spave{v}^{(n)}(\xvec_1,\ldots,\xvec_n)}
=  \int \dint{\xvec^*} d(\xvec^*)
\ave{v^{(n)}(\xvec_1,\ldots,\xvec_n;\xvec^*)}
\end{equation}
and $\ave{v^{(n)}(\xvec_1,\ldots,\xvec_n;\xvec^*)}$
denotes the $n$-point correlation function of 
the substrate particle density generated by a single diffusive particle 
started at $\xvec^*$. Equivalently in $\kvec$-space
\begin{equation*}
\ave{\spave{v}^{(n)}(\ldots)}
\!=\!
\int \ddint{x^*} d(\xvec^*)
\ave{v^{(n)}(\ldots;\xvec^*)}
\!=\!
\int \ddintbar{k^*} d(-\kvec^*) \ave{v^{(n)}(\ldots;\kvec^*)}.
\end{equation*}

Given that $\creat{b}{\xvec}\annih{b}{\xvec}$ is
the particle density operator, that correlation function is the
expectation
\begin{multline}
\ave{v^{(n)}(\xvec_1,\ldots,\xvec_n;\xvec^*)} 
= 
\lim_{t_1,t_2,\ldots,t_n\to\infty}
\left\langle
\psi^{\dagger}(\xvec_1,t_1)\psi(\xvec_1,t_1)
\right.
\\
\left.
\times\psi^{\dagger}(\xvec_2,t_2)\psi(\xvec_2,t_2)
\times \ldots \times
\psi^{\dagger}(\xvec_n,t_n)\psi(\xvec_n,t_n)
\times \phi^{\dagger}(\xvec^*,t_0)
\right\rangle
\end{multline}
with only a single,\footnote{Obviously, an entirely different set of
diagrams will be generated by having more than one initial particle,
clearly so in the presence of a finite carrying capacity, where the
trail of one suppresses the trail of the other, $\tikz[baseline=-3pt]{
\draw[tsubstrate](-0.5,0) -- (0,0); 
\draw[tactivity](0.7,0) -- (0,0);
\draw[tactivity](0.7,-0.2) -- (0.3,-0.2);
\draw[tsubstrate](0.3,-0.2) -- (0,0);
} \propto \sigma\tau$. One can see that this diagram
contains an infrared divergence for $d\ge4$
\cite{vandenBergBolthausendenHollander:2004}.} 
initial, diffusive particle started at $\xvec^*,t_0$. The multiple
limits on the right are needed so we measure deposition due to the
active particle left \emph{after} its lifetime. 
As the present
phenomenon is time-homogeneous, $t_0$ will not feature explicitly, but
rather enter in the difference $t_i-t_0$, each of which diverges as the
limits are taken. 
In principle, only a single
limit is needed, $t=t_1=t_2=\ldots=t_n\to\infty$, but as discussed
below, equal times leave some ambiguity that can be avoided.

For $n=1$, the relevant correlation function is 
$\ave{
\psi^{\dagger}(\xvec_1,t_1)\psi(\xvec_1,t_1)
\phi^{\dagger}(\xvec^*,0)}$, which leaves us with four terms after
replacing by Doi-shifted creation operators, 
\begin{multline}\elabel{v_1_term_by_term}
\ave{
\psi^{\dagger}(\xvec_1,t_1)\psi(\xvec_1,t_1)
\phi^{\dagger}(\xvec^*,0)}
\\
=
\ave{\psi(\xvec_1,t_1)}
+\ave{\psitilde(\xvec_1,t_1)\psi(\xvec_1,t_1)}\\
+\ave{\psi(\xvec_1,t_1)\phitilde(\xvec^*,0)}
+\ave{\psitilde(\xvec_1,t_1)\psi(\xvec_1,t_1)\phitilde(\xvec^*,0)} \ .
\end{multline}
Pure annihilation, $\ave{\psi}$, vanishes --- it is
the expected density of substrate particles in the vacuum, as no active
particle has been created first. The
expectation
$\ave{\psitilde(\xvec_1,t_1)\psi(\xvec_1,t_1)}\propto\theta(t_1-t_1)$ vanishes as
well, for $\theta(0)=0$ (effectively the It{\=o} interpretation
of the time derivatives, \cite{Taeuber:2014}) is needed in order to make the 
Doi-Pelitti approach meaningful. The field $\psitilde(\xvec_1,t_1)$
in the density $\psitilde(\xvec_1,t_1)\psi(\xvec_1,t_1)$ is meant to \emph{re}-create the particle annihilated by the operator
corresponding to $\psi(\xvec_1,t_1)$. For the same reason,
$\ave{\psitilde(\xvec_1,t_1)\psi(\xvec_1,t_1)\phitilde(\xvec^*,0)}$
vanishes, even when a vertex, 
\[
\tikz[baseline=-2.5pt]{
\draw[Aactivity] (0.7,0) -- (0,0) node[at start, right] {$0$};
\draw[substrate] (0,0) -- (-0.7,0) node[at end, left] {$t_1$};
\draw[substrate] (-50:0.7) -- (0,0) node[at start,right] {$t_1$};
}
\]
is available. In fact, to contribute,
any occurrence of 
$\psitilde(\xvec_1,t_1)$ requires an occurrence of $\psi(\xvec_2,t_2)$
with $t_2>t_1$. What remains of \Eref{v_1_term_by_term} is therefore 
only 
\begin{equation}\elabel{tree-level_V1}
\ave{\psi(\xvec_1,t_1)\phitilde(\xvec^*,0)}=\tikz[baseline=-2.5pt]{
\draw[activity] (0.6,0) -- (0,0);
\draw[substrate] (0,0) -- (-0.6,0);
}\ .
\end{equation}
Taking the Fourier transform 
of \Eref{deri_transmutation}, 
\begin{multline}\elabel{FT_deri_transmutation}
\int\dintbar{\omega_0}\exp{-\imag\omega_0 t_0}
\int\dintbar{\omega_1}\exp{-\imag\omega_1 t_1}
\frac{\deltabar(\kvec_0+\kvec_1)\deltabar(\omega_0+\omega_1)}{-\imag\omega_1+\epsilon'}
\ \tau \ \frac{1}{-\imag\omega_1+D\kvec_1^2+r}
\\
=
\frac{\deltabar(\kvec_0+\kvec_1) \theta(t_1-t_0)
\tau}{D\kvec_1^2+r-\epsilon'} \left(
\exp{-\epsilon'(t_1-t_0)} - \exp{-(D\kvec_1^2+r)(t_1-t_0)}
\right)
\end{multline}
reveals the general mechanism of
\begin{equation}\elabel{general_limt_mechanism}
\lim_{t\to\infty}\lim_{\epsilon'\to0}
\int\dintbar{\omega}
\frac{g(\omega)\exp{-\imag\omega t}}{-\imag\omega+\epsilon'}
=g(0) \ ,
\end{equation}
provided $g(\omega)$ itself has no pole at the origin, as otherwise
additional residues that survive the limit $t\to\infty$ would 
have to be considered.

In \Eref{FT_deri_transmutation}
the starting point of the walker still enters via $\kvec_0$. If that ``driving''
is done with a distribution of initial starting points $d(\kvec_0)$, the
resulting deposition is given by
\begin{equation}\elabel{tau_integral}
\ave{\spave{v}^{(1)}(\kvec)} 
=  \int \ddintbar{k_0} \ave{v^{(1)}(\kvec;\kvec_0)}\ d(-\kvec_0)
= \frac{\tau d(\kvec)}{D\kvec^2+r}
= 
\left.
\tikz[baseline=-2.5pt]{
\draw[Sactivity] (0.6,0) -- (0,0) node[at end,above] {$\tau$};
\draw[substrate] (0,0) -- (-0.6,0);
\draw[very thick,color=red,fill=white] (0.6,0) circle(0.1cm);
}\right|_{\omega=0}
\end{equation}
where the little circle on the right indicates the ``driving'' which ``supplies'' a
certain momentum distribution.
More specifically,  an initial
distribution of $d(\xvec)=\delta(\xvec-\xvec^*)$ has Fourier components 
$$\int\ddint{x} d(\xvec) \exp{-\imag \kvec_0 \xvec}
=\exp{-\imag \kvec_0 \xvec^*}
=d(\kvec_0)$$ and the 
resulting deposition is distributed according to
$$\ave{\spave{v}^{(1)}(\kvec;\xvec^*)}= \tau \exp{-\imag \kvec
\xvec^*}/(D\kvec^2+r)\ .$$
In an infinite system, the position of the initial driving should not
and will not enter --- to calculate the volume of the Sausage, we will
evaluate at $\kvec=0$. The same applies for the time of when the initial
distribution of particles is made. 
In principle it would give rise to an
additional factor of $\exp{-\imag \omega t^*}$, but we will evaluate at
$\omega=0$.

Evaluating at $\kvec=0$ in the bulk produces the volume integral over the offspring
distribution, \ie the expected volume $V$ of the Sausage, in the absence of a
limiting carrying capacity, 
\begin{equation}\elabel{V1_bulk}
\ave{V}=\ave{\spave{v}^{(1)}(\kvec=0;\xvec^*)}=\frac{\tau}{r}\ ,
\end{equation}
which corresponds to the
na{\"i}ve expectation of the (number) deposition rate $\tau$ multiplied by
the survival time of the random walker $1/r$. From this expression it is
also clear that the ``volume'' calculated here is, as expected, dimensionless.

Following similar arguments for $n=2$, the relevant diagrams are
\begin{multline}\elabel{second_order_diagrams}
\ave{
\psi^{\dagger}(\xvec_2,t_2)\psi(\xvec_2,t_2)
\psi^{\dagger}(\xvec_1,t_1)\psi(\xvec_1,t_1)
\phi^{\dagger}(\xvec_0,t_0)
} \\
=
\ave{
\psi(\xvec_2,t_2)
\psitilde(\xvec_1,t_1)\psi(\xvec_1,t_1)
\phitilde(\xvec_0,t_0)
} 
+
\ave{
\psi(\xvec_2,t_2)
\psi(\xvec_1,t_1)
\phitilde(\xvec_0,t_0)
} \\
\corresponding\ \tikz[baseline=-2.5pt]{
\draw[Sactivity] (0.6,0) -- (0,0) node[at end,above] {$\tau$};
\draw[substrate] (0,0) -- (-0.6,0);
\draw[substrate] (-0.8,0) -- (-1.4,0);
\crossblob{-0.7,0};
}
+
\tikz[baseline=-2.5pt]{
\draw[substrate] (-1.0,0) -- (-0.5,0) node[at end, above] {$\tau$};
\draw[Sactivity] (0,0) -- (-0.5,0);
\draw[Sactivity] (0.5,0) -- (0,0) node[at end, above] {$\sigma$};
\draw[substrate] (-130:0.7) -- (0,0);
} \ ,
\end{multline}
where the symbol $\tikz[baseline=-2.5pt]{\crossblob{0,0}}$ represents
$\psitilde(\xvec,t)\psi(\xvec,t)$, which is a convolution in Fourier
space, 
\begin{equation}
\tikz[baseline=-2.5pt]{\crossblob{0,0}}
\ \corresponding
\int\ddint{x}\dint{t} \psitilde(\xvec,t)\psi(\xvec,t) 
\exp{\imag\omega t-\imag\kvec \xvec}
=
\int\ddintbar{k'}\dintbar{\omega'}
\psitilde(\kvec',\omega')\psi(\kvec-\kvec',\omega-\omega')
\end{equation}
so that 
\begin{equation}\elabel{twiddle_blob_twiddle}
\tikz[baseline=-2.5pt]{
\draw[substrate] (0,0) -- (-1,0) node[at start,below] {$\kvec_0,\omega_0$};
\draw[substrate] (-1.2,0) -- (-2.2,0) node [at end, below] {$\kvec_2,\omega_2$};
\crossblob[$\kvec_1,\omega_1$]{-1.1,0};
} =
\frac{
\deltabar(\omega_0+\omega_1+\omega_2)
\deltabar(\kvec_0+\kvec_1+\kvec_2)
}{
(-\imag\omega_2 + \epsilon')
( \imag\omega_0 + \epsilon')
} \ ,
\end{equation}
which in real space and time gives a 
$\delta(\xvec_2-\xvec_1)\delta(\xvec_1-\xvec_0)\theta(t_2-t_1)\theta(t_1-t_0)$, corresponding to an
immobile particle deposited at $t_0$ and $\xvec_0$, found later at time $t_1>t_0$ and $\xvec_1=\xvec_0$
and left there to be found again at time $t_2>t_1$ and $\xvec_2=\xvec_1=\xvec_0$.

The effect of taking the limits $t_i\to\infty$ is the same as for the first
moment, namely it results in $\omega_i=0$. The same holds here, except that in
diagrams containing the convolution, the result depends on the order in which
the limits are taken. This can be seen in the factor
$\theta(t_2-t_1)\theta(t_1-t_0)$, as one naturally expects from this diagram:
The first probing must occur after creation and the second one after the
first. A diagram like the second in \Eref{second_order_diagrams} does not carry a constraint like that. 

Each of the diagrams on the right hand side of
\Eref{second_order_diagrams} appears twice, as the external fields can be
attached in two different ways. When evaluating at $\kvec_1=\kvec_2=0$ 
this would lead to the same (effective) combinatorial factor of $2$ for
both diagrams.
However, taking
the time limits in a particular order
means that one labelling of the first diagram results in a vanishing
contribution. The resulting combinatorial factors are 
therefore $1$ for 
$\tikz[baseline=-2.5pt,scale=0.7]{
\draw[tSactivity] (0.6,0) -- (0,0);
\draw[tsubstrate] (0,0) -- (-0.6,0);
\draw[tsubstrate] (-0.8,0) -- (-1.4,0);
\crossblob{-0.7,0};
}$ 
and $2$ for 
$\tikz[baseline=-2.5pt,scale=0.7]{
\draw[tsubstrate] (-1.0,0) -- (-0.5,0);
\draw[tSactivity] (0,0) -- (-0.5,0);
\draw[tSactivity] (0.5,0) -- (0,0);
\draw[tsubstrate] (-160:0.7) -- (0,0);
}$, \ie 
\begin{equation}\elabel{V2_bulk}
\ave{V^2}=\frac{\tau}{r}\left(1+2\frac{\sigma}{r}\right) \ ,
\end{equation}
again dimensionless. Given that $\tau=\sigma=\gamma$ initially,
\Eref{non-linearities}, the above may be written
$\gamma/r+2\gamma^2/r^2$. Unsurprisingly, the moments correspond to
those expected for a Poisson process with rate $\gamma$ taking place
during the exponentially distributed lifetime of the particle, subject
to a Poisson process with rate $r$. The resulting moment generating
function is simply 
\begin{equation}\elabel{MGF_tree_level}
\MC(x)=
\frac{r/\gamma}{r/\gamma+1-\exp{x}}
\end{equation}
with $\ave{V^n}=\left.\frac{\plaind^n}{\plaind x^n}\right|_{x=0} \MC(x)$
reproducing all moments once $\tau=\sigma=\gamma$.

Carrying on with the diagrammatic expansion,
higher order moments can be constructed correspondingly. At tree level (or
$n_0\to\infty$ equivalently), there are no further vertices contributing.
Determining $\ave{v^{(n)}(\kvec_1,\ldots,\kvec_n;\kvec_0)}$ is therefore merely a
matter of adding substrate legs, $\tikz[baseline=-2.5pt]{\draw[tsubstrate]
(0,0) -- (-0.5,0);}$, either by adding a convolution,
$\tikz[baseline=-2.5pt]{\crossblob{0,0}}$, or by
branching with coupling $\sigma$. For example, 
\begin{equation}\elabel{tree_v3}
\ave{v^{(3)}} 
\corresponding
\tikz[baseline=-2.5pt]{
\draw[Sactivity] (0.6,0) -- (0,0) node[at end,above] {$\tau$};
\draw[substrate] (0,0) -- (-0.6,0);
\draw[substrate] (-0.8,0) -- (-1.4,0);
\draw[substrate] (-1.6,0) -- (-2.2,0);
\crossblob{-0.7,0};
\crossblob{-1.5,0};
}
+
\tikz[baseline=-2.5pt]{
\draw[substrate] (-1.1,0) -- (-0.5,0) node[at end, above] {$\tau$};
\draw[Sactivity] (0,0) -- (-0.6,0);
\draw[substrate] (-1.9,0) -- (-1.3,0);
\crossblob{-1.2,0};
\draw[Sactivity] (0.5,0) -- (0,0) node[at end, above] {$\sigma$};
\draw[substrate] (-130:0.7) -- (0,0);
}
+
\tikz[baseline=-2.5pt]{
\draw[substrate] (-1.0,0) -- (-0.5,0) node[at end, above] {$\tau$};
\draw[Sactivity] (0,0) -- (-0.6,0);
\draw[Sactivity] (0.5,0) -- (0,0) node[at end, above] {$\sigma$};
\draw[substrate] (-130:0.6) -- (0,0);
\draw[substrate] (-130:1.4) -- (-130:0.8);
\crossblob{-130:0.7};
}
+
\tikz[baseline=-2.5pt]{
\draw[substrate] (-1.0,0) -- (-0.5,0) node[at end, above] {$\tau$};
\draw[Sactivity] (0,0) -- (-0.6,0);
\draw[Sactivity] (0.5,0) -- (0,0) node[at end, above] {$\sigma$};
\draw[substrate] (-130:0.6) -- (0,0);
\begin{scope}[xshift=0.5cm]
\draw[Sactivity] (0.5,0) -- (0,0) node[at end, above] {$\sigma$};
\draw[substrate] (-130:0.6) -- (0,0);
\end{scope}
}
.
\end{equation}
Upon taking the limits, effective combinatorial factors become $1$, $3$,
$3$ and $6$ respectively, so that 
\begin{equation}\elabel{V3_bulk}
\ave{V^3}=
\frac{\tau}{r} \left( 1 
 +  6 \frac{\sigma}{r}
 +  6 \left(\frac{\sigma}{r}\right)^2
\right) \ ,
\end{equation}
and similarly
\begin{subeqnarray}{\elabel{higher_bulk_moments}}
\ave{V^4}&=&\frac{\tau}{r} \left(
 1 
 +  14 \frac{\sigma}{r}
 +  36 \left(\frac{\sigma}{r}\right)^2
 +  24 \left(\frac{\sigma}{r}\right)^3
\right)\\
\ave{V^5}&=&\frac{\tau}{r} \left(
 1 
 +  30 \frac{\sigma}{r}
 +  150 \left(\frac{\sigma}{r}\right)^2
 +  240 \left(\frac{\sigma}{r}\right)^3
 +  120 \left(\frac{\sigma}{r}\right)^4
\right)\\
\ave{V^6}&=&\frac{\tau}{r} \left(
 1 
 +  62 \frac{\sigma}{r}
 +  540 \left(\frac{\sigma}{r}\right)^2
 +  1560 \left(\frac{\sigma}{r}\right)^3
 +  1800 \left(\frac{\sigma}{r}\right)^4
\right.\nonumber \\
&& \left. +  720 \left(\frac{\sigma}{r}\right)^5 \right) \ .
\end{subeqnarray}
In general, the leading order behaviour in small $r$ at tree level in the bulk is dominated
by diagrams with the largest number of branches, \ie the largest power of $\sigma$, like the 
right-most term in \Eref{tree_v3}, so that
\begin{equation}
\ave{V^m} \propto m! \tau\sigma^{m-1} r^{-m} \ ,
\elabel{general_scaling_tree}
\end{equation}
which is essentially determined by the time the active particle survives.

\subsection{Observables at tree level: open boundary conditions}\slabel{OBC}
Nothing changes diagrammatically when considering the observables
introduced above in systems with open boundary conditions along one axis. As 
$n_0\to\infty$ does not pose a constraint, it makes no difference
whether the system is periodically closed (in $d=2$ a finite cylinder) or
infinitely extended (infinite slab) along the
other
axes --- these directions simply do not matter for the observables
studied, except when the diffusion constant enters. What makes the difference to the considerations in the bulk,
\Sref{tree_level}, are \emph{open} dimensions, in the following fixed to
\emph{one}, so that the number of infinite (or, at this stage
equivalently, periodically closed) directions is $\dtilde=d-1$; in the following $\kvec,\kvec'\in\Rset^{\dtilde}$.

While the diagrams obviously remain unchanged, their interpretation
changes because of the orthogonality relations as stated in
\Eref{two_terms_Dirichlet} and 
\Eref{three_terms_Dirichlet} or, equivalently, the lack of momentum
conservation due to the absence of translational invariance.
Replacing the propagators by
\begin{subequations}\elabel{bare_propagators_open}
\begin{align}\elabel{bare_activity_propagator_open}
\ave{\phi_n(\kvec,\omega)\phitilde_m(\kvec',\omega')}_0
&=
\frac{\deltabar(\kvec+\kvec')\deltabar(\omega+\omega')\frac{L}{2}\delta_{n,m}}{-\imag\omega+D\kvec^2+Dq_n^2+r}
\ \corresponding\ \tikz[baseline=-2.5pt]{\draw[activity] (0.6,0) -- (-0.6,0);}\\
\ave{\psi_n(\kvec,\omega)\psitilde_m(\kvec',\omega')}_0
&=
\frac{\deltabar(\kvec+\kvec')\deltabar(\omega+\omega')\frac{L}{2}\delta_{n,m}}{-\imag\omega+\epsilon'}
\ \corresponding\ \tikz[baseline=-2.5pt]{\draw[substrate] (0.6,0) --
(-0.6,0);} \ ,
\end{align}
\end{subequations}
where a \emph{single} open dimension causes the appearance of the
indices $n$ and $m$,
results in the one point function
\[
\ave{\spave{v}^{(1)}_n(\kvec)} 
= \frac{\tau d_n(\kvec)}{D\kvec^2+Dq_n^2+r}
= 
\left.
\tikz[baseline=-2.5pt]{
\draw[Sactivity] (0.6,0) -- (0,0) node[at end,above] {$\tau$};
\draw[substrate] (0,0) -- (-0.6,0);
\draw[very thick,color=red,fill=white] (0.6,0) circle(0.1cm);
}\right|_{\omega=0} \ ,
\]
where the index $n$ refers to the Fourier-$\sin$ component as discussed
in \Sref{Fourier}. 
If driving (\ie initial deposition) is uniform (homogeneous) along the open, finite
axis, its Fourier transform is
$d_n(\kvec)=\deltabar(\kvec)\int_0^L\dint{z}\sin(q_nz)/L=2\deltabar(\kvec)/(q_nL)$ for odd $n$ and vanishes
otherwise. As for the periodic or infinite dimensions, the distribution
of the driving does not enter into $\ave{V^n}$, as momentum conservation
implies that the only amplitudes of the driving that matter are that of the
$\kvec=0$ or $k_0=0$ modes, \Eref{fourier_transform} and \Eref{efuncs_exp}.

Integrating $(2/L)\sum_n \ave{\spave{v}^{(1)}_n(\kvec)} \sin(q_nz)$ over
the interval $[0,L]$ produces 
\cite{Mathematica:8.0.1.0}
\begin{multline}\elabel{V_open_details}
\ave{V}=\frac{2}{L}\sum_{n\,\text{odd}} \frac{2}{q_n} 
\frac{\tau}{Dq_n^2+r} \frac{2}{Lq_n} \\
=
\frac{8\tau}{\pi^4 D}L^2
\sum_{n\,\text{odd}}
\frac{1}{n^2}\frac{1}{n^2+\frac{rL^2}{D\pi^2}}
=\frac{\tau}{r}\left(1-\sqrt{\frac{4D}{rL^2}}
\tanh\left( \sqrt{\frac{rL^2}{4D}}\right) \right) \ .
\end{multline}
In the limit of large $L$ this result recovers \Eref{V1_bulk}, which
would be less surprising if $L\to\infty$ would simply restore the bulk,
which is, however, not the case, because as the driving is uniform, some of it
always takes place ``close to'' the open boundaries. However, open
boundaries matter only up to a distance of $\sqrt{D/r}$ from the
boundaries, \ie the fraction of walkers affected by the open boundaries
is of the order $\sqrt{D/r}/L$.

The limit
$r\to0$ gives $\ave{V}=\tau L^2/(12 D)$, matching results for the
average residence time of a random walker on a finite lattice with
cylindrical boundary conditions using $D=1/(2d)$
\cite{Pruessner_aves:2013}. Sticking with $r\to0$,
calculating higher order moments for uniform driving is straight-forward, although
somewhat tedious. For example, the two diagrams contributing to
$\ave{v^{(2)}}$ are 
\begin{multline}\elabel{convolution_tree}
\tikz[baseline=-2.5pt]{
\draw[activity] (1,0) -- (0,0) node[at end, above] {$\tau$} node[at start,below] {$\kvec_0,\omega_0,n$};
\draw[substrate] (0,0) -- (-1,0);
\draw[substrate] (-1.2,0) -- (-2.2,0) node [at end, below] {$\kvec_2,\omega_2,l$};
\crossblob[$\kvec_1,\omega_1,m$]{-1.1,0};
} \\
=
\frac{
\tau \deltabar(\omega_0+\omega_1+\omega_2)
\deltabar(\kvec_0+\kvec_1+\kvec_2)
L \epsnml
}{
(-\imag\omega_2 + \epsilon')
( \imag\omega_0 + D\kvec_0^2 + Dq_n^2 +r)
( \imag\omega_0 + \epsilon')
}
\end{multline}
and
\begin{multline}\elabel{branch_tree}
\tikz[baseline=-2.5pt]{
\draw[substrate] (-1.8,0) -- (-0.9,0) node[at end, above] {$\tau$} node [at start, below] {$\kvec_2,\omega_2,l$}; 
\draw[Sactivity] (0,0) -- (-0.9,0);
\draw[Sactivity] (0.9,0) -- (0,0) node[at end, above] {$\sigma$} node[at start,below] {$\kvec_0,\omega_0,n$};
\draw[substrate] (-130:0.7) -- (0,0) node [at start, below] {$\kvec_1,\omega_1,m$};
} \\
=
\frac{
\tau \sigma \deltabar(\omega_0+\omega_1+\omega_2)
\deltabar(\kvec_0+\kvec_1+\kvec_2)
L \epsnml
}{
(-\imag\omega_2 + D\kvec_2^2 + Dq_{\ell}^2 +r)
(-\imag\omega_2 + \epsilon')
( \imag\omega_0 + D\kvec_0^2 + Dq_n^2 +r)
( \imag\omega_0 + \epsilon')
} \ .
\end{multline}
Using 
\begin{subeqnarray}{\elabel{messy_sums}}
2\pi \sum_{\substack{nm\ell\\\text{odd}}}
\frac{1}{n^3}
\frac{1}{m}
\frac{1}{\ell}
\epsnml &=& \frac{1}{6} \left(\frac{\pi}{2}\right)^6 \\
2\pi \sum_{\substack{nm\ell\\\text{odd}}}
\frac{1}{n^3}
\frac{1}{m}
\frac{1}{l^3}
\epsnml &=& \frac{1}{15} \left(\frac{\pi}{2}\right)^8 \ ,
\end{subeqnarray}
where $n,m,l\in\{1,3,5,\ldots\}$ (as driving is uniform and the sausage volume is an integral over the entire system),
then produces
\begin{equation}\elabel{V2_tree_open}
\ave{V^2}=\frac{\tau L^2}{12D} \left(1+\frac{\sigma L^2}{5D}\right) \ .
\end{equation}
This may be compared to the known expressions for the moments of the
number of distinct sites visited by a random walker within $n$ \emph{moves}
\cite[in particular Eq.~(A.14)]{Torney:1986}, which contains logarithms even in three dimensions, where the present tree level results are valid. This is, apparently,
caused by constraining the length of the Sausage by limiting the number
of moves, rather than a Poissonian death rate.

Performing the summations \Eref{messy_sums} is straight-forward, but
messy and tedious.\footnote{GP would like to thank Aman Pujara and 
Songhongyang Yuan for their help.} The relevant sums converge rather quickly,
for the third moment
producing (by summing numerically over $200$ terms for each index), for example
\begin{multline}\elabel{V3_tree_open}
\ave{V^3} 
=
\frac{\tau L^2}{12D}
\left(
1.00002196165\ldots+
0.60000307652\ldots \frac{\sigma L^2}{D}\right.\\
\left. + 0.060714286977\ldots \frac{\sigma^2 L^4}{D^2}
\right).
\end{multline}
Just like in the bulk for small $r$, \Eref{higher_bulk_moments},
the diagrams dominating large $L$ are the tree-branch-like diagrams such
as \Eref{branch_tree}, with highest power of $\sigma$, rather than those involving convolutions,
\Eref{convolution_tree}. Each new branch produces a factor $L^2$, so in
general
\begin{equation}\elabel{general_scaling_open_tree}
\ave{V^m}\propto \tau \sigma^{m-1} L^{2m} D^{-m}\ ,
\end{equation}
as in \Eref{general_scaling_tree} essentially determined by the time the
particle stays on the lattice.

Similar to the bulk, the lack of interaction allows the volume moments
of the Sausage to be determined on the basis of the underlying Poisson
process. In the case of homogeneous drive, the $m$th moment of the residence
time $t_r$ of a Brownian particle diffusing on an open interval of length $L$ is
\begin{equation}\elabel{exact_residence_moments}
\ave{t_r^m}=
\frac{8 m!}{\pi^{2(m+1)} D^m} L^{2m}\sum_{n\,\text{odd}} n^{-2(m+1)}
\end{equation}
and the moment generating function of the Poissonian deposition with
rate $\gamma$ is just $\MC(z)=\exp{-\gamma t_r (1-\exp{z})}$, so that
$\ave{V^m}=\ave{\plaind^m \MC(z)/\plaind z^m|{z=0}}$, reproducing the
results above such as
\begin{equation}
\ave{V^3}=\frac{\gamma L^2}{12D} 
\left(
1+
\frac{3 \gamma L^2}{5D}+
\frac{17 \gamma^2 L^4}{280 D^2}\ ,
\right)
\end{equation}
confirming, in particular, the high accuracy of the leading order term
in $L$, as $17/280=0.06071428571428571428\ldots$.

\section{Beyond tree level}\slabel{beyond_tree_level}
Below $d_c=2$ the additional vertices parameterised by $\lambda$, $\kappa$,
$\chi$ and $\xi$, \Eref{mu_kappa} and \Eref{rho_nu} respectively, have to
be taken into account. Because $\kappa$ is the only vertex that has the
same number of incoming and outgoing legs, it is immediately clear that
its presence can, and, in fact, will contribute to the renormalisation
of all other vertices, say
\begin{equation}\elabel{kappa_renormalises_sigma}
\tikz[baseline=-2.5pt]{
\draw[Aactivity] (0,0) -- (-0.4,0) node [at start,above] {$\sigma$};
\draw[Aactivity] (0.4,0) -- (0,0);
\draw[substrate] (-130:0.4) -- (0,0);
}
+
\tikz[baseline=-2.5pt]{
\draw[activity] (0.4,0) -- (-0.4,0) node [at start,above] {$\sigma$} node [at end,above] {$\kappa$};
\draw[Bsubstrate] (-0.4,0) to[bend right,in=-90,out=-90] (0.4,0);
\draw[Aactivity] (0.8,0) -- (0.4,0);
\begin{scope}[xshift=-0.4cm]
  \draw[Aactivity] (0,0) -- (-0.4,0);
  \draw[substrate] (-130:0.4) -- (0,0);
\end{scope}
}
+
\tikz[baseline=-2.5pt]{
\draw[activity] (0.4,0) -- (-0.4,0) node [at start,above] {$\sigma$} node [at end,above] {$\kappa$};
\draw[Bsubstrate] (-0.4,0) to[bend right,in=-90,out=-90] (0.4,0);
\draw[Aactivity] (0.8,0) -- (0.4,0);
\begin{scope}[xshift=-0.8cm]
\draw[Bsubstrate] (-0.4,0) to[bend right,in=-90,out=-90] (0.4,0);
\draw[activity] (0.4,0) -- (-0.4,0) node [at end, above] {$\kappa$};
\begin{scope}[xshift=-0.4cm]
  \draw[Aactivity] (0,0) -- (-0.4,0);
  \draw[substrate] (-130:0.4) -- (0,0);
\end{scope}
\end{scope}
}
+
\tikz[baseline=-2.5pt]{
\draw[activity] (0.4,0) -- (-0.4,0) node [at start,above] {$\sigma$} node [at end,above] {$\kappa$};
\draw[Bsubstrate] (-0.4,0) to[bend right,in=-90,out=-90] (0.4,0);
\draw[Aactivity] (0.8,0) -- (0.4,0);
\begin{scope}[xshift=-0.8cm]
\draw[Bsubstrate] (-0.4,0) to[bend right,in=-90,out=-90] (0.4,0);
\draw[activity] (0.4,0) -- (-0.4,0) node [at end, above] {$\kappa$};
\begin{scope}[xshift=-0.4cm]
  \draw[Aactivity] (0,0) -- (-0.4,0);
\end{scope}
\end{scope}
\begin{scope}[xshift=-1.6cm]
\draw[Bsubstrate] (-0.4,0) to[bend right,in=-90,out=-90] (0.4,0);
\draw[activity] (0.4,0) -- (-0.4,0) node [at end, above] {$\kappa$};
\begin{scope}[xshift=-0.4cm]
  \draw[Aactivity] (0,0) -- (-0.4,0);
  \draw[substrate] (-130:0.4) -- (0,0);
\end{scope}
\end{scope}
}
+ \ldots \ ,
\end{equation}
but in particular itself:
\begin{equation}\elabel{kappa_renormalises_itself}
\tikz[baseline=-2.5pt]{
\draw[Aactivity] (0,0) -- (-0.4,0) node [at start,above] {$\kappa$};
\draw[Aactivity] (0.4,0) -- (0,0);
\draw[substrate] (-130:0.4) -- (0,0);
\draw[substrate] (-50:0.4) -- (0,0);
}
+
\tikz[baseline=-2.5pt]{
\draw[activity] (0.4,0) -- (-0.4,0) node [at start,above] {$\kappa$} node [at end,above] {$\kappa$};
\draw[Bsubstrate] (-0.4,0) to[bend right,in=-90,out=-90] (0.4,0);
\draw[substrate,xshift=0.4cm] (-50:0.4) -- (0,0);
\draw[Aactivity] (0.8,0) -- (0.4,0);
\begin{scope}[xshift=-0.4cm]
  \draw[Aactivity] (0,0) -- (-0.4,0);
  \draw[substrate] (-130:0.4) -- (0,0);
\end{scope}
}
+
\tikz[baseline=-2.5pt]{
\draw[activity] (0.4,0) -- (-0.4,0) node [at start,above] {$\kappa$} node [at end,above] {$\kappa$};
\draw[Bsubstrate] (-0.4,0) to[bend right,in=-90,out=-90] (0.4,0);
\draw[substrate,xshift=0.4cm] (-50:0.4) -- (0,0);
\draw[Aactivity] (0.8,0) -- (0.4,0);
\begin{scope}[xshift=-0.8cm]
\draw[Bsubstrate] (-0.4,0) to[bend right,in=-90,out=-90] (0.4,0);
\draw[activity] (0.4,0) -- (-0.4,0) node [at end, above] {$\kappa$};
\begin{scope}[xshift=-0.4cm]
  \draw[Aactivity] (0,0) -- (-0.4,0);
  \draw[substrate] (-130:0.4) -- (0,0);
\end{scope}
\end{scope}
}
+
\tikz[baseline=-2.5pt]{
\draw[activity] (0.4,0) -- (-0.4,0) node [at start,above] {$\kappa$} node [at end,above] {$\kappa$};
\draw[Bsubstrate] (-0.4,0) to[bend right,in=-90,out=-90] (0.4,0);
\draw[substrate,xshift=0.4cm] (-50:0.4) -- (0,0);
\draw[Aactivity] (0.8,0) -- (0.4,0);
\begin{scope}[xshift=-0.8cm]
\draw[Bsubstrate] (-0.4,0) to[bend right,in=-90,out=-90] (0.4,0);
\draw[activity] (0.4,0) -- (-0.4,0) node [at end, above] {$\kappa$};
\begin{scope}[xshift=-0.4cm]
  \draw[Aactivity] (0,0) -- (-0.4,0);
\end{scope}
\end{scope}
\begin{scope}[xshift=-1.6cm]
\draw[Bsubstrate] (-0.4,0) to[bend right,in=-90,out=-90] (0.4,0);
\draw[activity] (0.4,0) -- (-0.4,0) node [at end, above] {$\kappa$};
\begin{scope}[xshift=-0.4cm]
  \draw[Aactivity] (0,0) -- (-0.4,0);
  \draw[substrate] (-130:0.4) -- (0,0);
\end{scope}
\end{scope}
}
+ \ldots \ .
\end{equation}

Among the vertices
introduced in \Sref{non-linearity}, namely
$\tau$
$\tikz[baseline=-2.5pt]{\draw[tsubstrate](0,0) -- (-0.4,0);\draw[tAactivity] (0,0) -- (0.4,0);}$, 
$\sigma$
$\tikz[baseline=-2.5pt]{\draw[tAactivity] (-0.4,0) -- (0.4,0);\draw[tsubstrate](-150:0.4) -- (0,0);}$, 
$\lambda$
$\tikz[baseline=-2.5pt]{\draw[tsubstrate](0,0) -- (-0.4,0);\draw[tAactivity] (0,0) -- (0.4,0);\draw[tsubstrate](-30:0.4) -- (0,0);}$, 
$\kappa$
$\tikz[baseline=-2.5pt]{\draw[tAactivity] (-0.4,0) -- (0.4,0);\draw[tsubstrate](-30:0.4) -- (0,0);\draw[tsubstrate](-150:0.4) -- (0,0);}$, 
$\chi$
$\tikz[baseline=-2.5pt]{\draw[tsubstrate](0,0) -- (-0.4,0);\draw[tAactivity] (0,0) -- (0.4,0);\draw[tsubstrate](160:0.4) -- (0,0);\draw[tsubstrate](-30:0.4) -- (0,0);}$ 
and
$\xi$
$\tikz[baseline=-2.5pt]{\draw[tAactivity] (-0.4,0) -- (0.4,0);\draw[tsubstrate](-30:0.4) -- (0,0);\draw[tsubstrate](-150:0.4) -- (0,0);\draw[tsubstrate](160:0.4) -- (0,0);}$, 
none has an outgoing activity leg if it
does not have an incoming activity leg, and all have at least as many
outgoing substrate legs as they have incoming substrate legs. Apart from
$\kappa$, each vertex has either more outgoing substrate legs than
incoming ones or fewer outgoing activity legs than incoming ones.
Combining them in any form will thus never result in a diagram
contributing to the renormalisation of $\kappa$, which has one leg of each kind.

Combinations of other vertices gives rise to ``cross-production'', say
$\chi$, 
$\tikz[baseline=-2.5pt]{
\draw[tAactivity] (0.5,0) -- (0,0);
\draw[tsubstrate] (-0.5,0) -- (0,0);
\draw[tsubstrate] (-20:0.5) -- (0,0);
\draw[tsubstrate] (160:0.5) -- (0,0);
}$, by $\lambda\xi$,
$\tikz[baseline=-2.5pt]{
\draw[tactivity] (1,0) -- (0,0);
\draw[tsubstrate] (0,0) -- (-0.5,0);
\draw[tsubstrate] (0,0) to[bend right,in=-140,out=-40] (1.,0);
\begin{scope}[xshift=1.cm]
\draw[tAactivity] (0.5,0) -- (0,0);
\draw[tAactivity] (0,0) -- (-0.5,0);
\draw[tsubstrate] (-20:0.5) -- (0,0);
\draw[tsubstrate] (160:0.5) -- (0,0);
\end{scope}
}$, but none of these terms contains more than one loop without the
involvement of $\kappa$. As for the generation of higher order vertices,
it is clear that the number of outgoing substrate-legs (on the left)
can never be decreased by combining vertices, because within every
vertex the number of outgoing substrate legs is at least that of
incoming substrate legs. In particular $\tikz[baseline=-2.5pt]{
\draw[tAactivity](0.4,0) -- (-0.4,0);
\draw[tsubstrate] (-20:0.4) -- (0,0);
}$
does not exist. A vertex like that, combined, say, with $\sigma$ to
form the bubble 
$\tikz[baseline=-2.5pt]{
\draw[tactivity] (0.8,0) -- (0.4,0);
\draw[tactivity] (-0.4,0) -- (-0.8,0);
\draw[tactivity] (0.4,0) to[bend left,in=-140,out=-40] (-0.4,0);
\draw[tsubstrate] (-0.4,0) to[bend right,in=-140,out=-40] (0.4,0);
}$, which renormalises the propagator, suggests the
diffusive movement of active particles is affected by the presence of
substrate particles. This is, by definition of the original problem, not
the case.

Because no active particles are generated solely by a combination of
substrate particles, none of the vertices has more outgoing then
incoming activity legs. Denoting the tree level coupling of the proper vertex (with amputated legs)
\begin{equation}
\Gammaai{m}{n}{a}{b} = 
\tikz[baseline=-2.5pt]{
\begin{scope}[rotate=0]
  \draw [decorate,decoration={brace,amplitude=5pt}] (-128:1.4cm) -- (-172:1.4) node[pos=0.5,left,xshift=-0.07cm,yshift=-0.14cm] {$a$};
  \draw[substrate] (-130:0.3) -- (-130:1.3);
  \path [postaction={decorate,decoration={raise=0ex,text along path,
  text align={center},
  text={|\large|....}}}]
  (-155:1.2cm) arc (-155:-130:1.2cm);
  \draw[substrate] (-155:0.3) -- (-155:1.3);
  \draw[substrate] (-170:0.3) -- (-170:1.3);
\end{scope}
\begin{scope}[rotate=-60]
  \draw [decorate,decoration={brace,amplitude=5pt}] (-128:1.4cm) -- (-172:1.4) node[pos=0.5,left,xshift=-0.07cm,yshift=0.14cm] {$m$};
  \draw[Aactivity] (-130:0.3) -- (-130:1.3);
  \path [postaction={decorate,decoration={raise=0ex,text along path,
  text align={center},
  text={|\large|....}}}]
  (-155:1.2cm) arc (-155:-130:1.2cm);
  \draw[Aactivity] (-155:0.3) -- (-155:1.3);
  \draw[Aactivity] (-170:0.3) -- (-170:1.3);
\end{scope}
\begin{scope}[rotate=120]
  \draw [decorate,decoration={brace,amplitude=5pt}] (-128:1.4cm) -- (-172:1.4) node[pos=0.5,right,xshift=0.07cm,yshift=-0.14cm] {$b$};
  \draw[substrate] (-130:0.3) -- (-130:1.3);
  \path [postaction={decorate,decoration={raise=0ex,text along path,
  text align={center},
  text={|\large|....}}}]
  (-155:1.2cm) arc (-155:-130:1.2cm);
  \draw[substrate] (-155:0.3) -- (-155:1.3);
  \draw[substrate] (-170:0.3) -- (-170:1.3);
\end{scope}
\begin{scope}[rotate=180]
  \draw [decorate,decoration={brace,amplitude=5pt}] (-128:1.4cm) -- (-172:1.4) node[pos=0.5,right,xshift=0.07cm,yshift=0.14cm] {$n$};
  \draw[Aactivity] (-130:0.3) -- (-130:1.3);
  \path [postaction={decorate,decoration={raise=0ex,text along path,
  text align={center},
  text={|\large|....}}}]
  (-155:1.2cm) arc (-155:-130:1.2cm);
  \draw[Aactivity] (-155:0.3) -- (-155:1.3);
  \draw[Aactivity] (-170:0.3) -- (-170:1.3);
\end{scope}
\draw[thick,fill=white] (0,0) circle (0.3cm) 
node {};
}
\end{equation}
of the correlation function
\begin{multline}
\Gai{m}{n}{a}{b}
\left(
\{\kvec_1,\ldots,\kvec_{m+n+a+b}; 
  \omega_1,\ldots,\omega_{m+n+a+b}\};
D,r,\tau,\sigma,\lambda,\kappa,\chi,\xi
\right)\\
=
\left\langle
\underbrace{\phi(\kvec_1,\omega_1)\ldots\phi(\kvec_m,\omega_m)}_{\textcolor{red}{m} \text{ terms}}
\underbrace{\psi\ldots\psi}_{\textcolor{olive}{a} \text{ terms}}
\underbrace{\phitilde\ldots\phitilde}_{\textcolor{red}{n} \text{ terms}}
\underbrace{\psitilde\ldots\psitilde}_{\textcolor{olive}{b} \text{ terms}}
\right\rangle
\elabel{general_G}
\end{multline}
by $\gammaai{m}{n}{a}{b}$, the topological conditions on the vertices
can be summarised as $a\ge b$, $b\le1$, $n=1$, $m\le1$, $m+a\ge 1$,
which means that there are in fact only four different types of
vertices, namely $n=1$, $b=0,1$ and $m=0,1$, whereas $a$ is hitherto
undetermined. For future reference, we note
\begin{subequations}
\elabel{tau_etc_as_gamma}
\begin{align}
\tau    & = \gammaai{0}{1}{1}{0}   & \sigma  & = \gammaai{1}{1}{1}{0}\\
\lambda & = \gammaai{0}{1}{1}{1}   & \kappa  & = \gammaai{1}{1}{1}{1}\\
\chi    & = \gammaai{0}{1}{2}{1}   & \xi     & = \gammaai{1}{1}{2}{1}
\end{align}
\end{subequations}

Dimensional analysis gives 
\[
\dimensionof{\gammaai{m}{n}{a}{b}}=\dimensionof{\Gammaai{m}{n}{a}{b}}=\Ldim^{d(n+b-1)}
\Tdim^{a-b-1} \Adim^{m-n+a-b} \Bdim^{n-m}\ .
\]
Because diffusion is to be maintained, it follows that $\Tdim=\Ldim^2$, yet, as indicated
above, the dimensions of $\Adim$ and $\Bdim$ are to some extent a matter of
choice. Leaving them undetermined results in $d(n+b-1)+2(a-b)\le2$ for
$\Gammaai{m}{n}{a}{b}$ to be relevant in $d$ dimensions. Setting, on the
other hand, $\Adim=\Bdim=\Tdim^{-1}$ (see above) results in $d(n+b-1)\le2$. As $n=1$, this
implies $(d-2)b+2a\le2$ and $db\le2$, respectively. In both cases, the
upper critical dimension for a vertex with $b\ge1$ and thus $a\ge1$ to
be relevant is $d_c=2$. On the other hand, no loop can be formed if
$b=0$, so above $d=2$ (where $b=1$ is irrelevant) there are no one-particle irreducibles
contributing to any of the $\Gammaai{m}{n}{a}{b}$ and so the set of
couplings introduced above, $\tau$, $\sigma$, $\lambda$, $\kappa$, $\chi$
and $\xi$ remains unchanged. As far as Sausage moments are concerned,
$\lambda$, $\kappa$, $\chi$ and $\xi$ do not enter, as there is no vertex
available to pair up the incoming substrate leg on the right. The
tree level results discussed in \Sref{tree_level} therefore are the
complete theory in $d>d_c=2$.

Below $d_c=2$, the dimensional analysis depends on the choice one makes
for $\Adim$ and $\Bdim$. If they remain independent, then the only relevant
vertices that are topologically possible are those with $a\le1$,
removing $\chi$ and $\xi$ from the problem. However, it is entirely
consistent (and one may argue, even necessary) to assume $\Adim=\Bdim=\Tdim^{-1}$,
resulting in no constraint on $a$ at all. Not only are therefore vertices
for all $a$ relevant, what is worse, they are all generated as
one-particle irreducibles. For example, the reducible
diagram 
$\tikz[baseline=-2.5pt]{
\draw[tsubstrate] (-1.0,0) -- (-0.5,0);
\draw[tSactivity] (0,0) -- (-0.5,0);
\draw[tSactivity] (0.5,0) -- (0,0);
\draw[tsubstrate] (-160:0.7) -- (0,0);
}$
contributing to $\ave{v^{(2)}}$ at tree level, \Sref{tree_level},
possesses, even
at one loop, two one-particle irreducible
counterparts in $d<2$,
\[
\tikz[baseline=-2.5pt]{
\draw[Sactivity] (0,0) -- (-1,0) node[at end, above] {$\ \ \chi$};
\draw[Sactivity] (0.5,0) -- (0,0) node[at end, above] {$\sigma$};
\draw[Bsubstrate] (-1,0) to[bend right,in=-90,out=-90] (0,0);
\draw[substrate] (-1,0)+(140:0.5) -- +(0,0);
\draw[substrate] (-1,0)+(220:0.5) -- +(0,0);
}
\qquad
\tikz[baseline=-2.5pt]{
\draw[Sactivity] (0,0) -- (-1,0) node[at end, above] {$\lambda$};
\draw[Sactivity] (0.5,0) -- (0,0) node[at end, above] {$\ \ \xi$};
\draw[Bsubstrate] (-1,0) to[bend right,in=-90,out=-90] (0,0);
\draw[substrate] (-1,0)+(180:0.5) -- +(0,0);
\draw[substrate] (-0,0)+(140:0.5) -- +(0,0);
}
\]
contributing to the corresponding proper vertex. Such diagrams exist for
all $a$, so, in principle, all these couplings have to be allowed for in
the Liouvillian and all have to be renormalised in their own right. The
good news is, however, 
that the $Z$-factor of $\kappa$ (see below) contains all infinities of all couplings exactly once, \ie the renormalisation of all couplings can be related to that of $\kappa$ by a 
diagrammatic vertex
identity, see \Sref{WardTakahashi}.

\subsection{Renormalisation}\slabel{renorm}
Without further ado, we will therefore carry on with renormalising $\kappa$
only. As suggested in \Eref{kappa_renormalises_itself}, this can be done
to all orders, in a geometric sum. The one and only
relevant integral is\footnote{We have written explicitly $\kappa$ vertices,
including the amputated legs. At this stage it is unimportant which
coupling forms the loop, but this will change when
we study infinite slabs in \Sref{semi_infty_strip}.}
\begin{multline}\elabel{one_loop}
\tikz[baseline=-2.5pt]{
\draw[Aactivity] (3,0) -- (2.4,0);
\draw[Aactivity] (0,0) -- (-0.6,0);
\draw[substrate] (0,0) -- (-120:0.7);
\draw[substrate] (2.4,0)+(0,0) -- +(-60:0.7);
\draw[activity] (2.4,0) -- (0,0) node[above,pos=0.5]
{$\kvec-\kvec',\omega-\omega'$} node[below,pos=0.5,yshift=-0.1cm] {$\kvec',\omega'$};
\draw[substrate] (0,0) to[bend right,in=-120,out=-60] (2.4,0);
} \\
= \kappa^2 \int \ddintbar{k'}\dintbar{\omega'}
\frac{1}{-\imag\omega'+\epsilon'}
\frac{1}{-\imag(\omega-\omega')+D(\kvec-\kvec')^2+r}\\
=
\frac{\kappa^2}{(4\pi)^{d/2}D}\left(\frac{r+\epsilon'-\imag\omega}{D}\right)^{-\epsilon/2}
\Gamma\left(\frac{\epsilon}{2}\right) \ ,
\end{multline}
where $\epsilon=2-d$ and we have indicated the \emph{total} momentum
$\kvec$ (\ie the sum of the momenta delivered by the two incoming legs) and the \emph{total} frequency $\omega$ going through
it.\footnote{Here and in the following we obviously choose
$\{\kvec_i,\omega_i\}=\{0,0\}$ in the renormalisation condition, \ie
$\gammaai{m}{n}{a}{b}_{\renorm} = \Gammaai{m}{n}{a}{b} (\{0,0\})$.}
This integral has the remarkable property that it is independent of $\kvec$,
because of the $\kvec$-independence of the substrate propagator. While
the latter conserves momentum in the bulk by virtue of 
$\deltabar(\kvec+\kvec')$ in \Eref{substrate_propagator}, its amplitude
does not depend on $\kvec$. Even if there were renormalisation of the
activity propagator it would therefore not affect its
$\kvec$-dependence, \ie $\eta=0$, whereas its $\omega$ dependence may be
affected, \ie $z\ne2$ would be possible. 

The expression $((r+\epsilon'-\imag\omega)/D)^{1/2}$ can be identified
as an inverse length; it is the infrared regularisation (or more precisely the normalisation point, $R=1$, \Eref{def_R}) that can, in the
present case, be implemented either by considering finite time
($\omega\ne0$), spontaneous extinction of activity ($r>0$) or, notably,
spontaneous extinction (evaporation) of substrate particles
($\epsilon'>0$). In order to extract exponents, it is replaced by
the arbitrary inverse length scale $\mu$. We will return to the case
$\mu=\sqrt{-\imag\omega/D}$ in \Sref{scaling}, \eg \Eref{V_t_exact}. For the time being,
the normalisation point is 
\begin{equation}\elabel{def_mu}
\mu^2=\frac{r}{D}
\end{equation}
with $\epsilon'\to0$, $\omega\to0$.

The renormalisation conditions are then (see \Eref{tau_etc_as_gamma})
\begin{subequations}
\elabel{tau_R_etc}
\begin{align}
\tau_{\renorm}    & = \Gammaai{0}{1}{1}{0}\left(\{0,0\}\right)   & \sigma_{\renorm}  & = \Gammaai{1}{1}{1}{0}\left(\{0,0\}\right)\\
\lambda_{\renorm} & = \Gammaai{0}{1}{1}{1}\left(\{0,0\}\right)   & \kappa_{\renorm}  & = \Gammaai{1}{1}{1}{1}\left(\{0,0\}\right)\\
\chi_{\renorm}    & = \Gammaai{0}{1}{2}{1}\left(\{0,0\}\right)   & \xi_{\renorm}     & = \Gammaai{1}{1}{2}{1}\left(\{0,0\}\right)
\end{align}
\end{subequations}
where $\{0,0\}$ indicates that the vertex is evaluated at vanishing momenta and frequencies. Defining $Z=\kappa_{\renorm}/\kappa$ allows all renormalisation to be expressed in terms of $Z$, as detailed in \Sref{WardTakahashi}.

Starting with only one loop,
the renormalisation of $\kappa$,
\Eref{kappa_renormalises_itself}, is therefore 
$\kappa_{\renorm} = \kappa - \kappa^2 W$ with
\begin{equation}
W = \frac{\Gamma\left(\frac{\epsilon}{2}\right)}{(4\pi)^{d/2}D}
\mu^{-\epsilon}
\elabel{def_W}
\end{equation}
or $\kappa_{\renorm}=\kappa Z$ with $Z=1-\kappa W$. 
Introducing the dimensionless
coupling $g=\kappa W/\Gamma(\epsilon/2)$ with $g_{\renorm}=gZ$ gives
$Z=1-g\Gamma(\epsilon/2)$, which may be approximated to one loop by
$Z=1-g_{\renorm}\Gamma(\epsilon/2)$.
Keeping, however, \emph{all} loops in \Eref{kappa_renormalises_itself},
this last expression is no longer an approximation:
if all terms in
\Eref{kappa_renormalises_itself} are retained, $Z$ becomes a
geometric sum in $g$,
\begin{equation}\elabel{Z_exact}
Z=1-\kappa W+(\kappa W)^2-\ldots=\frac{1}{1+\kappa W} =
\frac{1}{1+g\Gamma(\epsilon/2)} = 1-g_{\renorm}\Gamma(\epsilon/2) \ ,
\end{equation}
incorporating all parquet diagrams \cite{Herbut:2007}.
The resulting
$\beta$-function is 
$\beta_g(g) = \plaind g_{\renorm}/\plaind \ln \mu|_g g_{\renorm} = -\epsilon g_{\renorm} - \kappa W \beta_g$
and therefore
\begin{equation}
\beta_g(g) 
= \frac{-\epsilon g_{\renorm}}{1+\kappa W} 
= - \epsilon g_{\renorm} Z 
= -\epsilon g_{\renorm} \left(1-g_{\renorm} \Gamma(\epsilon/2)\right) \ .
\end{equation}
The last statement is exact to all orders; the non-trivial fixed point
in $\epsilon>0$ is exactly $g_{\renorm}^*=1/\Gamma(\epsilon/2)\approx\epsilon/2$, which is when the
$Z$-factor vanishes (as $g$ diverges in small $\mu$).

\subsubsection{Ward-Takahashi and vertex identities}\slabel{WardTakahashi}
Different vertices and therefore the renormalisation of different
couplings can be related to each other by Ward-Takahashi identities.
They are usually
constructed by considering global symmetries \cite{Zinn-Justin:1997}, such as
the invariance of the Liouvillian under \cite{BenitezWschebor:2013}
\begin{align}
\phi &\to \phi (1+\delta) & \phitilde &\to \phitilde (1+\delta)^{-1} \\
\psi &\to \psi (1+\delta) & \psitilde &\to \psitilde (1+\delta)^{-1} \ ,
\end{align}
to be considered for small $\delta$, 
which produces an identity on couplings involving an odd number of
fields,
\begin{equation}
(n-m+b-a) \Gammaai{m}{n}{a}{b}=
\int \ddint{x} \dint{t} \left(
-\sigma \frac{\partial}{\partial\sigma} 
+\lambda \frac{\partial}{\partial\lambda} 
-\xi \frac{\partial}{\partial\xi} 
\right) \Gammaai{m}{n}{a}{b} \ .
\elabel{WardTakahashi}
\end{equation}

The identities derived in the following are certainly consistent with
\Eref{WardTakahashi}, but derived at diagrammatic level. To start with,
we reiterate that \Eref{kappa_renormalises_itself} contains \emph{all} contributions (and to all
orders) to
$\Gammaai{1}{1}{1}{1}$, the renormalised vertex $\kappa$. 
Repeating 
for $\sigma$, 
$\tikz[baseline=-2.5pt]{\draw[tAactivity] (-0.4,0) -- (0.4,0);\draw[tsubstrate](-150:0.4) -- (0,0);}$, 
and $\lambda$,
$\tikz[baseline=-2.5pt]{\draw[tsubstrate](0,0) -- (-0.4,0);\draw[tAactivity] (0,0) -- (0.4,0);\draw[tsubstrate](-30:0.4) -- (0,0);}$, 
the diagrammatic, topological argument presented for $\kappa$ after
\Eref{kappa_renormalises_itself},
it turns out that diagrams contributing to their
renormalisation are essentially identical to those contributing to
$\kappa$, as shown in \Eref{kappa_renormalises_sigma}.
Using the same notation as in
\Eref{tau_R_etc}, we note that $\kappa_{\renorm}=\kappa Z$ implies
$\sigma_{\renorm}=\sigma Z$ and $\lambda_{\renorm}=\lambda Z$, \ie
\begin{equation}
\lambda_{\renorm}=\frac{\lambda}{\kappa}\kappa_{\renorm} \qquad
\sigma_{\renorm}=\frac{\sigma}{\kappa}\kappa_{\renorm} \ .
\end{equation}
The renormalisation of the coupling $\tau$ breaks with that pattern as
\begin{equation}\elabel{tau_R}
\tau_{\renorm} = \tau \left( 1 + \frac{\sigma\lambda}{\kappa \tau} (Z-1)\right) \ ,
\end{equation}
because the tree level contribution $\tau$, \Eref{deri_transmutation}, has higher
order
corrections such as
$\tikz[baseline=-2.5pt]{
\draw[tsubstrate] (-0.3,0) -- (0,0);
\draw[tAactivity] (1,0) -- (1.3,0);
\draw[tactivity] (1,0) -- (0,0);
\draw[tsubstrate] (0,0) to[bend right,in=-140,out=-40] (1.,0);
}$, which do not contain $\tau$ itself, but rather the combination
$\lambda\sigma$. However, at bare level, $\sigma=\tau$ and $\lambda=\kappa$, so that in the present case
\begin{equation}\elabel{tau_Ward_id}
\tau_{\renorm} = \frac{\tau}{\kappa} \kappa_{\renorm} \ .
\end{equation}

A different issue affects the renormalisation of $\chi$ and $\xi$. For
example, the
latter acquires contributions from any of the diagrams shown in
\Eref{kappa_renormalises_itself}
by ``growing an outgoing substrate leg'', $\tikz[baseline=-5.5pt]{
\draw[tsubstrate] (160:0.4) -- (0,0);}$, on any of the $\kappa$
vertices,
\begin{multline}\elabel{diagrams_xi}
\tikz[baseline=-2.5pt]{
\draw[substrate] (130:0.4) -- (0,0);
\draw[Aactivity] (0,0) -- (-0.4,0) node [at start,above] {$\ \xi$};
\draw[Aactivity] (0.4,0) -- (0,0);
\draw[substrate] (-130:0.4) -- (0,0);
\draw[substrate] (-50:0.4) -- (0,0);
}
+
\tikz[baseline=-2.5pt]{
\draw[substrate] (-0.4,0)+(130:0.4) -- +(0,0);
\draw[activity] (0.4,0) -- (-0.4,0) node [at start,above] {$\kappa$} node [at end,above] {$\ \xi$};
\draw[Bsubstrate] (-0.4,0) to[bend right,in=-90,out=-90] (0.4,0);
\draw[substrate,xshift=0.4cm] (-50:0.4) -- (0,0);
\draw[Aactivity] (0.8,0) -- (0.4,0);
\begin{scope}[xshift=-0.4cm]
  \draw[Aactivity] (0,0) -- (-0.4,0);
  \draw[substrate] (-130:0.4) -- (0,0);
\end{scope}
}
+
\tikz[baseline=-2.5pt]{
\draw[substrate] (0.4,0)+(130:0.4) -- +(0,0);
\draw[activity] (0.4,0) -- (-0.4,0) node [at start,above] {$\ \xi$} node [at end,above] {$\kappa$};
\draw[Bsubstrate] (-0.4,0) to[bend right,in=-90,out=-90] (0.4,0);
\draw[substrate,xshift=0.4cm] (-50:0.4) -- (0,0);
\draw[Aactivity] (0.8,0) -- (0.4,0);
\begin{scope}[xshift=-0.4cm]
  \draw[Aactivity] (0,0) -- (-0.4,0);
  \draw[substrate] (-130:0.4) -- (0,0);
\end{scope}
}
\\
+
\tikz[baseline=-2.5pt]{
\draw[activity] (0.4,0) -- (-0.4,0) node [at start,above] {$\kappa$} node [at end,above] {$\kappa$};
\draw[Bsubstrate] (-0.4,0) to[bend right,in=-90,out=-90] (0.4,0);
\draw[substrate,xshift=0.4cm] (-50:0.4) -- (0,0);
\draw[Aactivity] (0.8,0) -- (0.4,0);
\begin{scope}[xshift=-0.8cm]
\draw[Bsubstrate] (-0.4,0) to[bend right,in=-90,out=-90] (0.4,0);
\draw[substrate] (-0.4,0)+(130:0.4) -- +(0,0);
\draw[activity] (0.4,0) -- (-0.4,0) node [at end, above] {$\ \xi$};
\begin{scope}[xshift=-0.4cm]
  \draw[Aactivity] (0,0) -- (-0.4,0);
  \draw[substrate] (-130:0.4) -- (0,0);
\end{scope}
\end{scope}
}
+
\tikz[baseline=-2.5pt]{
\draw[substrate] (-0.4,0)+(130:0.4) -- +(0,0);
\draw[activity] (0.4,0) -- (-0.4,0) node [at start,above] {$\kappa$} node [at end,above] {$\ \xi$};
\draw[Bsubstrate] (-0.4,0) to[bend right,in=-90,out=-90] (0.4,0);
\draw[substrate,xshift=0.4cm] (-50:0.4) -- (0,0);
\draw[Aactivity] (0.8,0) -- (0.4,0);
\begin{scope}[xshift=-0.8cm]
\draw[Bsubstrate] (-0.4,0) to[bend right,in=-90,out=-90] (0.4,0);
\draw[activity] (0.4,0) -- (-0.4,0) node [at end, above] {$\kappa$};
\begin{scope}[xshift=-0.4cm]
  \draw[Aactivity] (0,0) -- (-0.4,0);
  \draw[substrate] (-130:0.4) -- (0,0);
\end{scope}
\end{scope}
}
+
\tikz[baseline=-2.5pt]{
\draw[substrate] (0.4,0)+(130:0.4) -- +(0,0);
\draw[activity] (0.4,0) -- (-0.4,0) node [at start,above] {$\ \xi$} node [at end,above] {$\kappa$};
\draw[Bsubstrate] (-0.4,0) to[bend right,in=-90,out=-90] (0.4,0);
\draw[substrate,xshift=0.4cm] (-50:0.4) -- (0,0);
\draw[Aactivity] (0.8,0) -- (0.4,0);
\begin{scope}[xshift=-0.8cm]
\draw[Bsubstrate] (-0.4,0) to[bend right,in=-90,out=-90] (0.4,0);
\draw[activity] (0.4,0) -- (-0.4,0) node [at end, above] {$\kappa$};
\begin{scope}[xshift=-0.4cm]
  \draw[Aactivity] (0,0) -- (-0.4,0);
  \draw[substrate] (-130:0.4) -- (0,0);
\end{scope}
\end{scope}
}
+ \ldots
\end{multline}
whereas contributions from 
$\tikz[baseline=-2.5pt]{
\draw[tactivity] (1,0) -- (0.5,0);
\draw[tactivity] (0.5,0) -- (0,0);
\draw[tsubstrate] (0,0) to[bend right,in=-140,out=-40] (1.,0);
\draw[tsubstrate] (0.5,0)+(160:0.5) -- +(0,0);
}$, generated by $\sigma \plaind/\plaind r$ are UV finite and therefore
dropped. Given that \Eref{diagrams_xi} are the only contributions to the
renormalisation of $\xi$, it reads
\begin{equation}
\xi_{\renorm}=2 \xi \frac{\plaind \kappa_{\renorm}}{\plaind \kappa} - \xi \frac{\kappa_{\renorm}}{\kappa}
\end{equation}
and correspondingly for the one-particle irreducible contributions to $\chi_{\renorm}$
\begin{equation}
\chi_{\renorm}=
2 \chi \frac{\plaind \kappa_{\renorm}}{\plaind \kappa} - \chi \frac{\kappa_{\renorm}}{\kappa}
\ ,
\end{equation}
where we have used $\chi-\xi\lambda/\kappa=0$.
From \Sref{renorm}, it is straight forward to show that 
\begin{equation}
\frac{\plaind \kappa_{\renorm}}{\plaind \kappa} = Z^2
\end{equation}
and we can therefore summarise
\begin{subequations}
\elabel{renorm_summary}
\begin{align}
\tau_{\renorm}    & = \tau Z     & \sigma_{\renorm}  & = \sigma Z \\
\lambda_{\renorm} & = \lambda Z  & \kappa_{\renorm}  & = \kappa Z \\
\chi_{\renorm}    & = \chi (2Z^2-Z)   & \xi_{\renorm}     & = \xi (2Z^2-Z)
\end{align}
\end{subequations}

In $d<2$, the only proper vertices $\Gammaai{n}{m}{a}{b}$ to consider are
those with $n=1$, $b\le1$, $m\le1$ and arbitrary $a$. The
renormalisation for all of them can be
traced back to that of $\Gammaai{1}{1}{1}{1}$. 
It is a matter of straight-forward algebra to demonstrate this explicitly. As these couplings play no further r{\^o}le for the observables analysed henceforth, we 
spare the reader a detailed account.

\subsection{Scaling}\slabel{scaling}
We are now in the position to determine the scaling of \emph{all}
couplings. For the time being, we will focus solely, however, on calculating the first moment of the Sausage volume. 

We have noted earlier (\Sref{beyond_tree_level}), that the governing
non-linearity is $\kappa$ and have already introduced the corresponding
dimensionless, renormalised coupling $g_{\renorm}$ and found its fixed point
value. Following the standard procedure \cite{Taeuber:2014}, we define the \emph{finite}, dimensionless,
renormalised vertex functions
\begin{multline}
\Gammaai{m}{n}{a}{b}\left(
\{\kvec, \omega\}; 
D, r, \tau, \sigma, \lambda, \kappa, \chi, \xi 
\right)\\
=
\mu^{2-d(n+b-1)} D
\GammaaiR{m}{n}{a}{b}\left(
\{\kvec, \omega\}; 
R, T_{\renorm}, s_{\renorm}, \ell_{\renorm}, g_{\renorm}, c_{\renorm}, x_{\renorm}; \mu
\right) \ ,
\elabel{def_GammaaiR}
\end{multline}
where $\{\kvec, \omega\}$ denotes the entire set of momenta and frequency arguments and $\mu$ is an arbitrary inverse scale.
In principle, there could be more bare couplings and there are certainly more generated, at least in principle, see \Sref{WardTakahashi}.
The vertex functions can
immediately be related to their arguments via \Eref{tau_R_etc} and \Eref{tau_etc_as_gamma}:
\begin{subequations}
\elabel{T_R_etc}
\begin{align}
R      & = r D^{-1} \mu^{-2} & \elabel{def_R}\\
T_{\renorm}    & = \tau Z D^{-1} \mu^{-2}                          & S_{\renorm}  & = \sigma Z D^{-1} \mu^{-2}\\
\ell_{\renorm} & = \lambda Z D^{-1} \mu^{-\epsilon} (4\pi)^{d/2}   & g_{\renorm}  & = \kappa Z D^{-1} \mu^{-\epsilon} (4\pi)^{d/2}\\
c_{\renorm}    & = \chi (2Z^2-Z) D^{-1} \mu^{-\epsilon}                 & x_{\renorm}  & = \xi (2Z^2-Z) D^{-1} \mu^{-\epsilon} \ ,
\end{align}
\end{subequations}
where the normalisation point is $R=1$. Because 
\[
\lim_{g_{\renorm}\to g_{\renorm}^*}
\frac{\plaind}{\plaind \ln \mu} \ln Z = \epsilon , 
\] 
$Z$ scales in $\mu$ like $Z\propto\mu^{\epsilon}$. 
The asymptotic solution (of the 
Callan-Symanzik equation) 
\begin{multline}
z^{2-d(n+b-1)} D\\
\times \GammaaiR{m}{n}{a}{b}\left(
\{\kvec, \omega\}; 
R z^{-2}, 
T_{\renorm} z^{-2+\epsilon}, 
s_{\renorm} z^{-2+\epsilon}, 
\ell_{\renorm}^*, 
g_{\renorm}^*, 
c_{\renorm}^*, 
x_{\renorm}^*; 
\mu z
\right) \\
= \text{asymptotically constant in small $z$}
\end{multline}
can be combined with 
the dimensional analysis of the renormalised vertex function, which gives
\begin{multline}
\GammaaiR{m}{n}{a}{b}\left(
\{\kvec, \omega\}; 
R, T_{\renorm}, s_{\renorm}, \ell_{\renorm}, g_{\renorm}, c_{\renorm}, x_{\renorm}; \mu z
\right)\\
=
\GammaaiR{m}{n}{a}{b}\left(
\left\{\frac{\kvec}{z}, \frac{\omega}{z^2}\right\}; 
R, T_{\renorm}, s_{\renorm}, \ell_{\renorm}, g_{\renorm}, c_{\renorm}, x_{\renorm}; \mu
\right) \ ,
\end{multline}
to give, using $z^2=r$ and \Eref{def_GammaaiR}, 
\begin{multline}
\Gammaai{m}{n}{a}{b}\left(
\{\kvec, \omega\}; 
D, r, \tau, \sigma, \lambda, \kappa, \chi, \xi 
\right)\\
=
r^{1-\frac{d}{2}(n+b-1)}
\Gammaai{m}{n}{a}{b}\left(
\{\frac{\kvec}{\sqrt{r}}, \frac{\omega}{r}\}; 
D, 1, 
\tau r^{-\frac{2-\epsilon}{2}}, 
\sigma r^{-\frac{2-\epsilon}{2}}, 
\lambda, \kappa, 
\chi, 
\xi
\right)
\end{multline}
As far as scaling (but not amplitudes) is concerned,
the tree level results apply to the right hand side as its mass $r$ is finite, \ie
\begin{equation}
\Gammaai{1}{1}{0}{0}\left(\{0, 0\}; D, r, \tau, \sigma, \lambda, \kappa, \chi, \xi \right)
= r^1 \times 1
\end{equation}
and
\begin{equation}
\Gammaai{0}{1}{1}{0}\left(\{0, 0\}; D, r, \tau, \sigma, \lambda, \kappa, \chi, \xi \right)
\propto r^1 \times \tau r^{-\frac{2-\epsilon}{2}}
\end{equation}
so that following \Eref{V1_bulk}
\begin{equation}\elabel{V_bulk_r}
\ave{V}
=
\frac{\Gammaai{0}{1}{1}{0}}{\Gammaai{1}{1}{0}{0}}
\propto
\tau r^{-d/2} \ .
\end{equation}
If $r^{-1}$ is interpreted as the observation time $t$, the result
\begin{equation}\elabel{V_bulk_t}
\ave{V} \propto t^{d/2}
\end{equation} 
in $d<2$ (and $\ave{V} \propto t$ in $d>2$,
\Eref{V1_bulk}) recovers the earlier result in \cite{BerezhkovskiiMakhnovskiiSuris:1989},
including the logarithmic corrections expected at the upper critical
dimension. Eqs.~\eref{V_bulk_r} and \eref{V_bulk_t} are the first two key results for the field theory of the Wiener Sausage reported in the present work. We will now further explore the results and their implications.

\begin{figure}
\begin{center}
\begin{tikzpicture}
\draw[->] (-5,0) -- (5,0) node[at end, below] {$x$};
\draw[line width=0.3cm] (-3,0.3cm) -- (3,0.3cm);
\draw[color=white,ultra thick] (-3,0.3cm) -- (3,0.3cm);
\draw (-3.8,0.3) -- (3.8,0.3);
\draw [|<->|] (3,0.7) -- (3.8,0.7) node[pos=0.5,above] {$\frac{V_0}{2}$};
\draw (3.8,0.4)+(-0.1,0) arc (90:-90:0.1);
\draw [|<->|] (-3,0.7) -- (-3.8,0.7) node[pos=0.5,above] {$\frac{V_0}{2}$};
\draw (-3.8,0.2)+(0.1,0) arc (90:-90:-0.1);
\draw [|<->|] (-3,0.7) -- (3,0.7) node[pos=0.5, above] {points actually visited};
\end{tikzpicture}
\end{center}
\caption{\flabel{1D_footprint}The volume of the Wiener Sausage in one dimension is the length
covered by the Brownian particle (the set of all points actually
visited) plus the volume $V_0$ of the sphere the Brownian
particle is dragging (indicated by the two rounded bumpers).}
\end{figure}

In $d=1$, it is an exercise in complex analysis (albeit lengthy) to determine the amplitude of the first moment. To make contact with established results in the literature, we study the sausage in one dimension after finite time $t$. Following the tree level results Eqs.~\eref{tree-level_V1}, \eref{tau_integral} and \eref{V1_bulk} we now have
\begin{equation}\elabel{V1_integral}
\ave{V}(t) = \int \dint{x_1} 
\tikz[baseline=-2.5pt]{
\draw[activity] (0.8,0) -- (0.2,0);
\draw[substrate] (-0.2,0) -- (-0.8,0);
\draw[thick,fill=white] (0,0) circle (0.2cm);
}
= \int \dintbar{\omega} \frac{1}{-\imag \omega + \epsilon'} \tau Z \frac{1}{-\imag \omega + r} \ ,
\end{equation}
where the space integral is taken by setting $\kvec=0$ and the driving has been evaluated to $d(0)=1$, see \Eref{tau_integral}. 
The $Z$-factor is given by \Eref{Z_exact}, but $\mu$ should be replaced
by $\sqrt{-\imag\omega/D}$, as we will consider the double limit
$r,\epsilon'\to0$, but at finite $\omega$, which is the \emph{total}
frequency flowing through the diagram, \Eref{one_loop},
so for $d=1=\epsilon$
\begin{equation}\elabel{Z_exact_d_1}
Z=
\frac{1}{1+\kappa \sqrt{\imag / (4 D \omega)}}
\end{equation}
which for small $\omega$ and therefore large $t$ (which we are interested in) is dominated by $2 \sqrt{-\imag D \omega)}/\kappa$. 
Keeping only that term, the integral in \Eref{V1_integral} can be performed and gives
\begin{equation}\elabel{V_t_exact}
\ave{V}(t) = \frac{\tau}{\kappa} 4 \sqrt{\frac{tD}{\pi}} \ .
\end{equation}
On the lattice, \ie before taking the continuum limit, sites have no volume and
the ratio $\tau/\kappa$ is just the carrying capacity $n_0$. Setting that to
unity one recovers, up to the additive volume mentioned above, see \Fref{1D_footprint},
the result in the continuum by 
Berezhkovskii, Makhnovskii and Suris 
\cite[Eq.~(10)]{BerezhkovskiiMakhnovskiiSuris:1989} which coincides with the 
asymptote on the lattice \cite{Montroll:1964,Torney:1986}.
Given the difference in the process and the course a field-theoretic
treatment taken, in particular the continuum limit, one might argue that this is a mere
coincidence. In fact, attempting a similar calculation for the amplitude of the second
moment does not suggest that it can be recovered.

As for higher moments of the volume, in addition to the two diagrams mentioned in \Eref{second_order_diagrams}, there is now also
\begin{equation}
\tikz[baseline=-2.5pt]{
\draw[substrate] (-130:1.3) -- (-130:0.6);
\draw[substrate] ( 130:1.3) -- ( 130:0.6);
\draw[Sactivity] (1.3,0) -- (0.6,0);
\draw[thick,fill=white] (0,0) circle (0.6cm) node {$\Gammaai{0}{1}{2}{0}$};
}
\end{equation}
and $\Gammaai{0}{1}{2}{0}= \chi\sigma(3Z-2Z^2-1)/\kappa$. However, as
above, the second moment is dominated in small $r$ by the second, tree-like term in
\Eref{second_order_diagrams}, which gives to leading order
\begin{equation}\elabel{V2_bulk_r}
\ave{V^2} \propto 2 \frac{\tau Z \sigma Z }{r^2} \propto 2 \tau \sigma r^{-d} \ ,
\end{equation}
as $Z\propto r^{\epsilon/2}$. Higher order moments follow that pattern
$\ave{V^m}\propto Z^m$,
and as
dimensional consistency is maintained by the dimensionless product
$r D^{d/\epsilon} \kappa^{-2/\epsilon}$ entering $Z$, Eqs.~\eref{def_mu}, \eref{def_W} and \eref{Z_exact}, 
\[
Z=\frac{1}{1+\Gamma\left(\frac{\epsilon}{2}\right) (4\pi)^{-d/2} (r D^{d/\epsilon} \kappa^{-2/\epsilon})^{-\epsilon/2}}
\]
the general result is
\begin{equation}\elabel{general_scaling}
\ave{V^m} \propto 
m! \tau \sigma^{m-1} r^{-m} \left( \frac{r D^{d/\epsilon}}{\kappa^{2/\epsilon}} \right)^{\epsilon m/2} 
=
m! \tau \sigma^{m-1} r^{-md/2} \left( \frac{D^{d/2}}{\kappa} \right)^{m} \ ,
\end{equation}
for $d<2$ with $r\corresponding 1/t$. Compared to
\Eref{general_scaling_tree} the diffusion constant is present again, as
the coverage depends not only on the survival time (determined by $r$),
but also on the area explored during that time.

\subsection{Infinite slab}\slabel{semi_infty_strip}
In the following, we study the renormalisation of the present field
theory on an infinite slab, \ie a lattice that is finite and open (Dirichlet
boundary conditions) along \emph{one} axis and infinite in $\dtilde=d-1$ orthogonal
dimensions. The same setup was considered at tree level in \Sref{OBC}.
Again, there
are no 
diagrammatic changes, yet the renormalisation procedure itself requires
closer attention. 

Before carrying out the integration of the relevant loop,
\Eref{one_loop}, we make a mild adjustment with respect to the set of
orthogonal functions that we use for the substrate and the activity.
While the latter is subject to Dirichlet boundary conditions in the
present case, naturally leading to the set of $\sin(q_n z)$
eigenfunctions introduced above, the former is not afflicted with such a
constraint, \ie in principle one may choose whatever set is most
convenient\footnote{The existence of a $0$-mode as the space integral is
one feature to consider.} and suitable.  As general as that statement is, there are,
however, some subtle implications; to start with, whatever
representation is used in the harmonic part of the Hamiltonian must
result in the integrand factorising, so that the path integral over the
Gaussian can be performed. In the presence of transmutation, that
couples the choice of the set for one species to that for the other.
With a suitable choice, all propagators fulfil orthogonality relations
and therefore conserve momentum, \ie they are proportional to
$\delta_{n,m}$ (in case of the basis $\sin(q_nz)$), 
$\delta_{n,-m}$ (basis $\exp{\imag k_nz}$)
and/or
$\delta(\kvec+\kvec')$ (basis $\exp{\imag \kvec z}$), which is obviously a welcome
simplification of the diagrams and their corresponding integrals and
sums. 

This constraint 
can be relaxed by considering transmutation
only perturbatively, \ie removing it from the harmonic part. However, if
different eigenfunctions are chosen for different species, transmutation
vertices are no longer momentum conserving; if we choose, as we will
below, $\sin(q_n z)$ for the basis of the activity and $\exp{\imag k_m z}$,
then the proper vertex of $\tau$ comes with 
\begin{equation}\elabel{def_Delta}
\int_0^L \dint{z} \exp{\imag k_n z} \sin(q_m z) =
L \Delta_{n,m}
\end{equation}
and a summation of the $n$ and $m$, connecting from the sides,
\Eref{deri_transmutation}, \ie
\begin{multline}\elabel{full_tau_open}
\tikz[baseline=-2.5pt]{
\draw[activity] (1,0) -- (0,0) node[at end,above] {$\tau$} 
  node[at start, below] {$\ell$} node [at end, below, xshift=0.2cm] {$m$};
\draw[substrate] (0,0) -- (-1,0) node [at start, below, xshift=-0.2cm] {$n$}
  node[at end, below] {$p$};
} 
\corresponding\ \frac{2}{L}\sum_{n=1}^\infty \frac{1}{L} \sum_{m=-\infty}^\infty
\int \ddMOneintbar{\!k}\, \dintbar{\omega''}\\
\times \ \ave{\psi_p(\kvec,\omega)\psitilde_n(\kvec'',\omega'')}
\ \tau L \Delta_{n,m}
\ave{\phi_m(\kvec'',\omega'')\phitilde_{\ell}(\kvec',\omega')}\\
=
\frac{\deltabar(\kvec+\kvec')\deltabar(\omega+\omega')}{-\imag\omega+\epsilon'}
\ \tau L \Delta_{p,\ell} \ \frac{1}{-\imag\omega+D\kvec^2+D q_{\ell}^2+r}
\end{multline}
where the $m\in\Zset$ refers to the index of the eigenfunction used for
the activity and $n\in\Nset^+$ to the eigenfunction of the substrate field. The fact that $\Delta_{p,l}$ has
off-diagonal elements indicates that momentum-conservation is broken.
Obviously, in the presence of boundaries, translational invariance is 
always broken, but that does not necessarily result in a lack of
momentum conservation in bare propagators, as it does here. However, it
always results in a lack of momentum conservation in vertices with more
than two legs, as only exponential eigenfunctions have the property that their products
are eigenfunctions as well. If propagators renormalise through these
vertices, they will eventually inherit the non-conservation, \ie
allowing them to have off-diagonal elements from the start will become a necessity in the process of renormalisation.

While the transmutation vertex introduced above may appear unnecessarily messy, it does not renormalise and does not require much further attention.
Rewriting the four-point vertex $\kappa$ in terms of the two different
sets of eigenfunctions, however, proves beneficial. Introducing
\begin{equation}\elabel{def_U}
\int_0^L \dint{z} 
\sin(q_n z) \exp{\imag k_m z} \exp{\imag k_k z} \sin(q_{\ell} z)
=
L U_{n,m+k,\ell}
\end{equation}
means that the relevant loop is 
\begin{multline}\elabel{one_loop_open}
\tikz[baseline=-2.5pt]{
\draw[Aactivity] (3,0) -- (2.4,0) node[at start, above,xshift=0.1cm] {$\ell$};
\draw[Aactivity] (0,0) -- (-0.6,0) node[at end, above,xshift=-0.1cm] {$n$};
\draw[substrate] (0,0) -- (-120:0.7) node[at end,
above,xshift=-0.3cm,yshift=-0.2cm] {$m_2$};
\draw[substrate] (2.4,0)+(0,0) -- +(-60:0.7) node[at end,
above,xshift=0.3cm,yshift=-0.2cm] {$m_1$};
\draw[activity] (2.4,0) -- (0,0) 
node[above,pos=0.5] {$\kvec\!-\!\kvec',\omega\!-\!\omega',n'$} 
node[below,pos=0.5,yshift=-0.07cm] {$\,\kvec'\!,\omega'\!,m'$};
\draw[substrate] (0,0) to[bend right,in=-120,out=-60] (2.4,0);
} \\
= 
\kappa^2 \frac{2}{L^2}\sum_{m'=-\infty}^{\infty} \sum_{n'=1}^{\infty}
\int \ddMOneintbar{k'}\dintbar{\omega'}
L^2 
U_{n,m_2-m',n'}
U_{n',m'+m_1,\ell}\\
\times
\frac{1}{-\imag\omega'+\epsilon'}
\frac{1}{-\imag(\omega-\omega')+D(\kvec-\kvec')^2+D q_{n'}^2+r} \ .
\end{multline}
Contrary to \Eref{one_loop}, it is now of great importance to know with
which couplings (here two $\kappa$ couplings)
this loop was formed, because different couplings require different
``tensors'', like $U_{n,m+k,\ell}$ in the present case. For example, the coupling $\sigma$ comes
with 
$\int_0^L \dint{z} \sin(q_n z) \exp{\imag k_m z} \sin(q_{\ell} z)$. The
actual technical difficulty to overcome, however, is the possible
renormalisation of $U_{n,m,\ell}$ itself, as there is no guarantee that the
right hand side of \Eref{one_loop_open} is proportional to $U_{n,m,\ell}$.
In other words, the sum \Eref{kappa_renormalises_itself} may be of the
form $\kappa (L U_{n,m+k,\ell} + \kappa W L U'_{n,m+k,\ell} + \kappa^2 W^2
L U''_{n,m+k,\ell}+\ldots)$, with $U'_{n,m+k,\ell}\ne U'_{n,m+k,\ell}$ \etc, rather than $L U_{n,m+k,\ell} \kappa (1+\kappa W +
\kappa^2 W^2 + \ldots)$, which would spoil the renormalisation process. 

Carrying on with that in mind, the integrals over $\omega'$ and $\kvec'$ are identical to the ones
carried out in \Eref{one_loop} and therefore straight-forward. The
summation over $m'$ is equally simple, because that index features only
in $U_{n,m,\ell}$ and \Eref{exp_sum} implies 
\begin{multline}\elabel{UU_integral}
\frac{1}{L}
\sum_{m'} 
L^2
U_{n,m_2-m',n'}
U_{n',m'+m_1,\ell}
\\
=
\int_0^L \dint{z}
\sin(q_n z)
\exp{\imag k_{m_2} z}
\exp{\imag k_{m_1} z}
\sin(q_{\ell} z)
\sin^2(q_{n'} z) \ .
\end{multline}

Using that identity in \Eref{one_loop_open} allows us to write
\begin{multline}\elabel{one_loop_open_next_step}
\tikz[baseline=-2.5pt]{
\draw[Aactivity] (1.25,0) -- (1.,0);
\draw[Aactivity] (0,0) -- (-0.25,0);
\draw[substrate] (0,0) -- (-120:0.3);
\draw[substrate] (1,0)+(0,0) -- +(-60:0.3);
\draw[Sactivity] (1,0) -- (0,0);
\draw[substrate] (0,0) to[bend right,in=-120,out=-60] (1.,0);
} =
\frac{\kappa^2}{(4\pi D)^{\frac{d-1}{2}}} 
\Gamma\left(\frac{3-d}{2}\right)\\
\times \int_0^L \dint{z}
\sin(q_n z)
\exp{\imag k_{m_2} z}
\exp{\imag k_{m_1} z}
\sin(q_{\ell} z)\\
\times
\frac{2}{L}\sum_{n'=1}^{\infty}
\left(D q_{n'}^2 + r + \epsilon' -\imag \omega\right)^{\frac{d-3}{2}}
\sin^2(q_{n'} z) \ .
\end{multline}
It is only that last sum that requires further investigation. In
particular, if we were able to demonstrate that it is essentially
independent of $z$, then the preceding integral becomes 
$L U_{n,m_1+m_2,\ell}$ and this contribution to the renormalisation of
$\kappa U_{n,m_1+m_2,\ell}$ is proportional to $U_{n,m_1+m_2,\ell}$. 

\newcommand{\Li}{\operatorname{Li}}

The remaining summation in \Eref{one_loop_open_next_step}
can be performed \cite{Mathematica:8.0.1.0} to leading order in the small\footnote{\fnlabel{open_recovers_bulk}For $\rho$
large, $\sum_{n'=1}^{\infty} (n'^2 + \rho)^{\frac{d-3}{2}} \approx 
\rho^{\frac{d-2}{2}} \sqrt{\pi}\Gamma(\epsilon/2)/\Gamma((3-d)/2)$, the
open system recovers the results in the bulk, \Eref{one_loop}.}
dimensionless quantity
$\rho=L^2 \left(r + \epsilon' -\imag \omega\right)/(\pi^2 D)$, 
\begin{multline}\elabel{open_sum}
\sum_{n'=1}^{\infty} (n'^2 + \rho)^{\frac{d-3}{2}} \sin^2(q_{n'} z) \\
= 
\half \zeta(3-d) - \quarter 
\left(
\Li_{3-d}\left(\exp{\frac{2\pi \imag z}{L}}\right)
+
\Li_{3-d}\left(\exp{-\frac{2\pi \imag z}{L}}\right)
\right) + \OC(\rho)
\end{multline}
with $\zeta(3-d)=\zeta(1+\epsilon)=(1/\epsilon) + \OC(1)$, the Riemann
$\zeta$-function, and $\Li_s(z)$ the polylogarithm with \cite{Mathematica:8.0.1.0}
\begin{equation}\elabel{polylogarithm}
\Li_{1+\epsilon}\left(\exp{\frac{2\pi \imag z}{L}}\right)
+
\Li_{1+\epsilon}\left(\exp{-\frac{2\pi \imag z}{L}}\right)
=-\ln(4 \sin^2(z\pi/L))+\OC(\epsilon) \ ,
\end{equation}
so that the leading order behaviour in $\epsilon$ of
\Eref{one_loop_open_next_step} is in fact
\begin{equation}\elabel{one_loop_open_final_step}
\tikz[baseline=-2.5pt]{
\draw[Aactivity] (1.25,0) -- (1.,0);
\draw[Aactivity] (0,0) -- (-0.25,0);
\draw[substrate] (0,0) -- (-120:0.3);
\draw[substrate] (1,0)+(0,0) -- +(-60:0.3);
\draw[Sactivity] (1,0) -- (0,0);
\draw[substrate] (0,0) to[bend right,in=-120,out=-60] (1.,0);
} =
\frac{2\kappa^2 L U_{n,m_1+m_2,\ell}}{(4\pi)^{d/2}D} 
\left(\frac{L}{\pi}\right)^{\epsilon} 
\zeta(3-d) + \OC(\epsilon^0,\rho)
\end{equation}
to leading order in $\epsilon$, where we have used $\Gamma((3-d)/2)=\sqrt{\pi}+\OC(\epsilon)$,
anticipating no singularities around $d=3$.

Approximating $2\zeta(3-d)\approx\Gamma(\epsilon/2)$ 
the $Z$-factor for the renormalisation of $\kappa$ in a system with open
boundaries in one dimension is therefore unchanged, \cf Eqs.~\eref{one_loop} and \eref{one_loop_open_final_step}, provided
$\mu=\pi/L$. Of course, that result holds only as long as 
$\rho\ll 1$ is small enough, in particular $r\ll D/L^2$, \ie sudden
death by extinction is rare compared to death by reaching the boundary.
In the case of more frequent deaths by extinction, or, equivalently,
taking the thermodynamic limit in the finite, open dimension, extinction is
expected to take over eventually and the bulk results above apply,
\Sref{scaling}. Although there is an effective change of mechanism (bulk
extinction versus reaching the edge), there is no dimensional crossover.

The renormalisation of $\tau$ involves the $\kappa$-loops characterised
above, as well as $\sigma$ and $\lambda$, which, in principle, have to
be considered separately; after all, the loop they form has a structure,
$\tikz[baseline=-2.5pt]{
\draw[Aactivity] (1.25,0) -- (1.,0);
\draw[substrate] (0,0) -- (-0.25,0);
\draw[Sactivity] (1,0) -- (0,0);
\draw[substrate] (0,0) to[bend right,in=-120,out=-60] (1.,0);
}$,
that deviates from the structure studied above, 
$\tikz[baseline=-2.5pt]{
\draw[Aactivity] (1.25,0) -- (1.,0);
\draw[Aactivity] (0,0) -- (-0.25,0);
\draw[substrate] (0,0) -- (-120:0.3);
\draw[substrate] (1,0)+(0,0) -- +(-60:0.3);
\draw[Sactivity] (1,0) -- (0,0);
\draw[substrate] (0,0) to[bend right,in=-120,out=-60] (1.,0);
}$,
\Eref{one_loop_open_final_step}. 
In principle, there is (again) no guarantee that the diagrams
contributing to the renormalisation of $\tau$ all have the same dependence
on the external indices, \ie whether they are all proportional to
$\Delta_{n,m}$, \Eref{def_Delta}. By definition, however, \Eref{def_U}
\begin{equation}\elabel{U2Delta}
\frac{2}{L}\sum_{n=1,3,\ldots}^{\infty} L U_{n,m,\ell}\frac{2}{q_n} =
L \Delta_{m,l} \ ,
\end{equation}
\ie one leg is removed by evaluating at $m_1=0$ (see the diagram in \Eref{one_loop_open}) and one by performing the
summation. Applying this operation to all diagrams appearing in
\Eref{kappa_renormalises_itself} produces all diagrams renormalising 
$\tau$. Provided that $\sigma=\tau$ and $\lambda=\kappa$, the renormalisation of $\tau$ is therefore linear in that of $\kappa$ and \Eref{tau_Ward_id} remains valid, \ie 
the renormalisation procedure outlined above for $\tau$ and
$\kappa$ remains intact. 

In principle, further attention is required for
the renormalisation of higher order vertices, but as long as only (external)
substrate legs are attached,
$\tikz[baseline=-2.5pt]{\draw[tsubstrate] (0,0) -- (-0.5,0);}$,
their index $m_n$ can be absorbed into the sum of the indices of the
substrate legs present: Just like any external leg can take up momentum or
frequency, such new legs shift the index used in the internal summation
such as the one in \Eref{one_loop_open}, but that does not affect the
renormalisation provided that it is done at vanishing external momenta, so that the external momenta do not move the poles of the propagators involved. 

We conclude that all 
diagrammatic vertex
identities of \Sref{WardTakahashi} remain unchanged. As for
the scaling of the Sausage volume, comparing
\Eref{one_loop_open_final_step} to \Eref{one_loop} and identifying
$\mu=\pi/L$ or $r=\pi^2 D/L^2$ means that now
\begin{equation}\elabel{general_scaling_open}
\ave{V^m} \propto 
m! \tau \sigma^{m-1} 
\left(\frac{L}{\pi}\right)^{md}
\kappa^{-m}
\end{equation}
for $d<2$, 
compared to \Eref{general_scaling}. Noticeably, compared to the tree
level \Eref{general_scaling_open_tree}, the diffusion constant is absent ---
in dimensions $d<2$ each point is visited infinitely often, regardless
of the diffusion constant. Even though the deposition in the present
setup is Poissonian, what determines the volume of the sausage is not the
time it takes the active
particles to drop off the lattice, $\propto L^2/D$, but the competition between
deposition parameterised by $\tau$ and $\sigma$ and its inhibition by $\kappa$.

The scaling $\ave{V^m} \propto L^{md}$ for $d<2$ suggests that the Wiener Sausage
is a ``compact'' $d$ dimensional object in dimensions $d<2$, whereas
$\ave{V^m} \propto L^{2m}$ at tree level, $d>2$, \Sref{OBC}. The Wiener
Sausage may therefore be seen as a two-dimensional object projected into
a $d$-dimensional space.

The obvious interpretation of
$r=\pi^2 D/L^2$ in \Eref{general_scaling_open} is that of $\pi/L$ being the lowest mode in the denominator of
the propagator \Eref{bare_activity_propagator_open} in the presence of open boundaries compared to
(effectively) $\sqrt{r/D}$ at $\kvec=0$ in
\Eref{bare_activity_propagator}.

It is interesting to determine the amplitude of the scaling in $L$ with
one open boundary, not least in order to determine whether the finding
of \Eref{V_t_exact} being identical to the result known in the literature is a mere
coincidence. Technically, the route to take differs from
\Eref{V_open_details}, because in \Sref{OBC} both substrate as well as
activity were represented in the $\sin$ eigensystem. However,
integrating over $L$ (for uniform driving and in order to determine the volume)
amounts to evaluating the matrix $\Delta_{p,\ell}$ in
\Eref{full_tau_open} at $p=0$ and in that case $L \Delta_{p,\ell}=2/q_{\ell}$
for $\ell$ odd and $0$ otherwise, which reproduces \Eref{V_open_details} at
$r=0$, with $\tau$ replaced by $\tau_{\renorm}$:
\begin{equation}
\ave{V}=\frac{2}{L}\sum_{n\,\text{odd}} \frac{2}{q_n} 
\frac{\tau_{\renorm}}{Dq_n^2} \frac{2}{Lq_n}
=
\frac{8\tau_{\renorm}}{\pi^4 D}L^2
\sum_{n\,\text{odd}}
\frac{1}{n^4}
=
\frac{\tau_{\renorm}}{12 D}L^2 \ .
\end{equation}
To determine $\tau_{\renorm}=\tau Z$ we replace $W$ in Eqs.~\eref{one_loop}, \eref{def_W} and \eref{Z_exact} by
$2 (L/\pi)^{\epsilon} \zeta(3-d)\Gamma((3-d)/2)/(\sqrt{\pi}D(4\pi)^{d/2})$, according to
\Eref{one_loop_open_final_step}, so that asymptotically in large $L$
\begin{equation}\elabel{V_open_L}
\ave{V}=\frac{\pi^{(5-d)/2}2^d \tau}{24  \zeta(3-d)
\Gamma\left(\frac{3-d}{2}\right)
\kappa}L^d
\end{equation}
which for $d=1$ reproduces the exact result (for uniform driving)
\begin{equation}\elabel{V_open_exact}
\ave{V}=\frac{ \tau}{2 \kappa}L \ ,
\end{equation}
which is easily confirmed from first principles. 
However, repeating the calculation for driving at the centre, $x^*=L/2$,
gives $d_n=(-1)^{(n-1)/2}$ for $n$ odd and $0$ otherwise, so that in 
$d=1$ after some algebra
\begin{equation}\elabel{V_open_centre}
\ave{V}(x^*=L/2)=\frac{3 \tau}{4 \kappa}L \ ,
\end{equation}
which is somewhat off the exact amplitude of $\ln(2)=0.69314718\ldots$
compared to $3/4$. This is apparently due to the renormalisation of $U_{n,m,\ell}$ in \Eref{one_loop_open_final_step} being correct only up to $\OC(\epsilon^0)$, but that problem may require further investigation.

\subsection{Infinite cylinder: crossover}\slabel{infinite_cylinder}
At tree level, \Sref{OBC}, it makes no technical difference to study the Sausage on
a finite cylinder or an infinite slab, because the relevant
observables require integration in space which amounts to evaluating at
$k_n=0$ or $\kvec=0$ resulting in the same expression, \eg
\Eref{V1_bulk} in both cases.

When including interaction, however, it does matter whether the lattice
studied is infinite in $d-1$ dimensions or periodically closed. 
Clearly a periodically closed axis has a $0$-mode and does therefore not impose an effective cutoff in $\kvec$.
In that respect, periodic closure is identical to infinite extent, while physically it is not (just like at tree level). 
One may therefore wonder how periodic closure differs from infinite extent mathematically: How does a finite cylinder differ from an infinite strip?
As a
first step to assess the effect, we \emph{replace} the open dimension (axis) by
a periodically closed one. One may regard this as an unfortunate kludge
--- after all, what we are really interested in is a system that is
finite in two dimensions, namely open in one and periodically closed in the
other. However, if the aim is to study finite size scaling in
$2-\epsilon$ dimensions, then two finite dimensions are already
$\epsilon$ too many.

However, the setup of an infinitely long (in $d-1$ dimensions)
periodically closed tube with circumference $L$ does address the problem in question, namely the difference of $\kvec=0$ in an infinitely extended axis versus $k_n=0$ in a finite but periodic closed dimension. In addition, an infinite cylinder compared to an infinite strip has
translational invariance restored in the periodic dimension, and therefore
the vertices even for a finite system dramatically simplified.

The physics of a $d$-dimensional system with one
axis periodically closed is quite clear: At early times, or,
equivalently, large extinction rates $r\gg D/L^2$, the periodic closure is
invisible and so the scaling is that of a $d$-dimensional (infinite)
bulk system as described in \Sref{scaling}, $\ave{V^m}\propto
r^{-md/2}$. But when the walker starts to re-encounter, due to the
periodic closure, sites visited earlier, this ``dimension will
saturate'' and so for very small $r$, it will display the scaling of
an infinite $d-1$-dimensional lattice. 

Just like for the setup in \Sref{tree_level}, it is most convenient to
study the system for small but finite extinction rate $r$. The integrals
to be performed are identical to \Eref{one_loop_open}, but both sums
have a pre-factor of $1/L$, \Eref{convenient_choices_periodic}, (rather than one having $1/L$ and
the other $2/L$, \Eref{convenient_choices}) and $L U_{n,m,l}$ has the much simpler
Kronecker form
\begin{equation}\elabel{def_Utilde}
\int_0^L \dint{z} 
\exp{\imag k_n z} \exp{\imag k_m z} 
\exp{\imag k_k z} \exp{\imag k_{\ell} z}
=
L \tilde{U}_{n,m+k,\ell}
= L \delta_{n+m+k+\ell,0} \ .
\end{equation}
Most importantly the expression corresponding to \Eref{UU_integral}
sees
$\sin^2(q_{n'}z)$ replaced by unity, because the bare propagator
corresponding to \Eref{bare_activity_propagator_open} carries a factor
$L\delta_{n+m,0}$, \Eref{efuncs_exp}, rather than
$L\delta_{n,m}/2$, \Eref{efuncs_sin}, which 
results in $n'$ of $\tilde{U}_{n,m_2-m',n'}$ to pair up with $-n'$ in
$\tilde{U}_{-n',m'+m_1,\ell}$. 
For easier comparison, we will keep $L\tilde{U}_{n,m+k,\ell}$ in the following.
We thus have
(see \Eref{one_loop_open_next_step})
\begin{multline}\elabel{one_loop_cyl_next_step}
\tikz[baseline=-2.5pt]{
\draw[Aactivity] (1.25,0) -- (1.,0);
\draw[Aactivity] (0,0) -- (-0.25,0);
\draw[substrate] (0,0) -- (-120:0.3);
\draw[substrate] (1,0)+(0,0) -- +(-60:0.3);
\draw[Sactivity] (1,0) -- (0,0);
\draw[substrate] (0,0) to[bend right,in=-120,out=-60] (1.,0);
}\\ =
\frac{\kappa^2 L \tilde{U}_{n,m+k,\ell}}{(4\pi D)^{\frac{d-1}{2}}} 
\Gamma\left(\frac{3-d}{2}\right) 
\frac{1}{L}\sum_{n'=-\infty}^{\infty}
\left(D k_{n'}^2 + r + \epsilon' -\imag \omega\right)^{\frac{d-3}{2}} \ .
\end{multline}
Comparing \Eref{one_loop_cyl_next_step} to \Eref{one_loop_open_next_step}, \Eref{open_sum} and
\Eref{one_loop_open_final_step} and re-arranging terms gives for small
$\rhotilde=L^2 \left(r + \epsilon' -\imag \omega\right)/(4 \pi^2 D)$
\begin{multline}\elabel{one_loop_cyl_final_step}
\tikz[baseline=-2.5pt]{
\draw[Aactivity] (1.25,0) -- (1.,0);
\draw[Aactivity] (0,0) -- (-0.25,0);
\draw[substrate] (0,0) -- (-120:0.3);
\draw[substrate] (1,0)+(0,0) -- +(-60:0.3);
\draw[Sactivity] (1,0) -- (0,0);
\draw[substrate] (0,0) to[bend right,in=-120,out=-60] (1.,0);
} =
\frac{2\kappa^2 L \tilde{U}_{n,m_1+m_2,\ell}}{(4\pi)^{d/2} D}
L^{\epsilon}
\Gamma\left(\frac{3-d}{2}\right)\\
\times
\left\{
(2\pi)^{-\epsilon} 
\frac{\zeta(3-d)}{\sqrt{\pi}}
+
\sqrt{\pi} 
\left(
\frac{r+\epsilon'-\imag \omega}{D}L^2
\right)^{\frac{d-3}{2}}
\right\} 
+ \OC(\epsilon^0,\rhotilde)\\
\end{multline}
and for large $\rhotilde$ 
\begin{multline}\elabel{one_loop_cyl_final_step_large_circ}
\tikz[baseline=-2.5pt]{
\draw[Aactivity] (1.25,0) -- (1.,0);
\draw[Aactivity] (0,0) -- (-0.25,0);
\draw[substrate] (0,0) -- (-120:0.3);
\draw[substrate] (1,0)+(0,0) -- +(-60:0.3);
\draw[Sactivity] (1,0) -- (0,0);
\draw[substrate] (0,0) to[bend right,in=-120,out=-60] (1.,0);
} =
\frac{\kappa^2 L \tilde{U}_{n,m_1+m_2,\ell}}{(4\pi)^{d/2} D}
\left(\frac{r+\epsilon'-\imag\omega}{D}\right)^{-\frac{\epsilon}{2}}
\Gamma\left(\frac{\epsilon}{2}\right)
+ \OC(\epsilon^0,\rhotilde^{\frac{d-3}{2}})
\end{multline}
using 
\begin{subnumcases}{\elabel{cyl_sum}
\hspace*{-1cm}
\sum_{n'=-\infty}^{\infty} (n'^2 + \rhotilde)^{\frac{d-3}{2}} =}
\rhotilde^{\frac{d-3}{2}} + 2 \zeta(3-d) + \OC(\rhotilde) &
for $\rhotilde \ll 1$ \elabel{cyl_sum_small}\\
\rhotilde^{\frac{d-2}{2}} 
\frac{\sqrt{\pi} \Gamma\left(\frac{2-d}{2}\right)}{\Gamma\left(\frac{3-d}{2}\right)} 
+ \OC\left(\rhotilde^{\frac{d-3}{2}}\right)&
for $\rhotilde \gg 1$ \elabel{cyl_sum_large} \ .
\end{subnumcases}
The asymptotics above are responsible for all the interesting features
to be discussed in the following. Firstly, intuition seems to play
tricks: One may think that for small $\rhotilde$ in the sum on the left of \Eref{cyl_sum}, it will always be
large compared to $n'=0$ and always be small compared to $n'\to\infty$.
In fact, one might think there is no difference at all between large or
small $\rhotilde$ and be tempted to approximate the sum immediately by an
integral, $\sum (n'^2 + \rhotilde)^{(d-3)/2}\approx
\int_{-\infty}^{\infty} \dint{n'} (n'^2 + \rhotilde)^{(d-3)/2} = 
\rhotilde^{\frac{d-2}{2}} \sqrt{\pi}
\Gamma\left(\frac{2-d}{2}\right)/\Gamma\left(\frac{3-d}{2}\right)$.
That, however, produces only the second line, \Eref{cyl_sum_large}.
The crucial difference is that in a sum each summand actually
contributes, whereas in an integral the integrand is weighted by the
integration mesh. So, the summand $(n'^2 + \rhotilde)^{(d-3)/2}$ has to
be evaluated for $n'=0$, producing $\rhotilde^{\frac{d-3}{2}}$ in
\Eref{cyl_sum_small}, which dominates the sum for $d<2$ (even $d<3$, but
the series does not converge for $2<d$, and, in fact, is not needed as
no IR divergences appear in
$d>2$) and
$\rhotilde\to0$. The remaining terms can actually be evaluated for
$\rhotilde=0$, producing $2\zeta(3-d)$. The integral, which the (Riemann) sum converges to for large $\rhotilde$, on the other hand,
is strictly proportional to $\rhotilde^{\frac{d-2}{2}}$ and therefore
much less divergent than then sum for small $\rhotilde\to0$ and $d<2$.

Of the two regimes $\rhotilde\gg1$ and $\rhotilde\ll1$ the former is more
easily analysed. Setting $\epsilon'-\imag\omega=0$ for the time being,
we notice that $\rhotilde\propto L^2 r$ suggests, somewhat
counter-intuitively, that large $r$, which shortens the lifetime of the
walker, has the same effect as large $L$, which prolongs the time it
takes the walker to explore the system. Both effects are, however, of
the same nature: They prevent the walker from ``feeling'' the
periodicity of the system. In that case, the walker displays bulk
behaviour and in fact, \Eref{one_loop_cyl_final_step_large_circ} is the same as
\Eref{one_loop}.

The other regime, $\rhotilde\ll1$ is richer. At $d<2$ and fixed $L$,
\Eref{one_loop_cyl_final_step} displays a
crossover between the two additive terms on the right hand
side. Stretching the expansion \eref{cyl_sum_small} beyond its stated limits, for
intermediate values of $r$ or $L$, $\rhotilde\approx 1$, the first
term on the right hand side of \Eref{one_loop_cyl_final_step} dominates
and the scaling
behaviour is that of an \emph{open} infinite slab of linear extent
$L$, \Eref{one_loop_open_final_step}. This is because at moderately
large $r$ (or, equally, short times $t$), the walker is not able to
fully explore the infinitely extended directions. But rather than ``falling off'' as
in the system with open boundaries, it starts crossing its
own path due to the periodic boundary conditions, at which point the
scaling like a $d$-dimensional bulk lattice ($\rhotilde\gg1$) ceases and turns into that of a $d$-dimensional open one ($\rhotilde\approx1$). The crossover can
also be seen in \Eref{cyl_sum_small}, which for $d<2$ is dominated by $2\zeta(3-d)$ for
large $\rhotilde$ and by $\rho^{(d-3)/2}$ for small
$\rhotilde$.

As $r$ gets even
smaller (or $t$ increases), $\rhotilde\to0$, the scaling is dominated by the infinite
dimensions, of which there are $\dtilde=d-1$, \ie the scaling is that of
a bulk system with $\dtilde$ dimensions as discussed in \Sref{renorm}, in
particular \Eref{one_loop}. In this setting, the walker explores an
infinitely long,
thin cylinder, which has effectively degenerated into an infinitely long
line. While the (comparatively) small circumference of the cylinder
remains accessible this is fully explored very quickly compared to the
progress in the infinite directions.

To emphasise the scaling of the last two regimes,
one can
re-write 
\Eref{one_loop_cyl_final_step} as
\begin{multline}\elabel{one_loop_cyl_final_step_simplified}
\tikz[baseline=-2.5pt]{
\draw[Aactivity] (1.25,0) -- (1.,0);
\draw[Aactivity] (0,0) -- (-0.25,0);
\draw[substrate] (0,0) -- (-120:0.3);
\draw[substrate] (1,0)+(0,0) -- +(-60:0.3);
\draw[Sactivity] (1,0) -- (0,0);
\draw[substrate] (0,0) to[bend right,in=-120,out=-60] (1.,0);
} =
\frac{2 \kappa^2 L \tilde{U}_{n,m_1+m_2,\ell}}{(4\pi)^{d/2}D} 
\frac{\Gamma\left(\frac{1+\epsilon}{2}\right)}{\sqrt{\pi}}
\left(\frac{L}{2\pi}\right)^{\epsilon} 
\zeta(1+\epsilon) \\
+
\frac{\kappa^2 L \tilde{U}_{n,m_1+m_2,\ell}}{(4\pi)^{\dtilde/2}LD}
\left(
\frac{r+\epsilon'-\imag \omega}{D}
\right)^{-\epsilontilde/2}
\Gamma\left(\frac{\epsilontilde}{2}\right)
+
\OC(\epsilon^0,\rhotilde) \ ,
\end{multline}
with 
$\epsilontilde=1+\epsilon=3-d$, 
$\dtilde=d-1$. 
Here, the first term displays the behaviour of the infinite slab
discussed above (\Sref{semi_infty_strip},
\Eref{one_loop_open_final_step}, $\zeta(3-d)\propto1/\epsilon$, but $L/\pi$ there and $L/(2\pi)$ here) and the second term that of a
bulk-system with $\dtilde$ dimensions, \Eref{one_loop}; the infrared singularity
$(r+\epsilon'-\imag \omega)^{-\epsilontilde/2}$ is in fact accompanied
by the corresponding ultraviolet singularity $\Gamma(\epsilontilde/2)$,
exactly as if the space dimension was reduced from $d$ to $\dtilde=d-1$.

The second term also
reveals an additional factor $1/L$ 
compared to \eref{one_loop}.\footnote{The factor $L$ in front of
$\Utilde_{n,m_1+m_2,\ell}$ should be regarded as part of the latter, as
the tree level obviously comes with the same pre-factor, see \Eref{def_Utilde}. 
Dimensional consistency is maintained by $((r+\epsilon'-\imag\omega)/D)^{-\epsilontilde/2}$, which has dimension $\Ldim^{1+\epsilon}$.} This expression
determines the factor $W$, which enters the $Z$-factor inversely,
$Z\propto L r^{\epsilontilde/2}$,
\Eref{Z_exact}, \ie in the present setting, the Sausage volume scales
like $(\tau/r) L r^{\epsilontilde/2} = \tau L r^{-\dtilde/2}$. The
scaling in $t$ is found by replacing $r$ by
$1/t$, or more precisely by $\omega$
and Fourier transforming according to \Eref{V1_integral}, which results
in the scaling
$\ave{V}\propto L t^{1-\epsilontilde/2}=Lt^{\dtilde/2}$.

\section{Summary and discussion}\slabel{discussion}
Because the basic process analysed above is very well understood and has
a long-standing history \cite{KolmogoroffLeontowitsch:1933,Spitzer:1964,KacLuttinger:1974,DonskerVaradhan:1975,LeGall:1986,BerezhkovskiiMakhnovskiiSuris:1989,vanWijlandCaserHilhorst:1997,Sznitman:1998,vandenBergBolthausendenHollander:2001,Spitzer:2001}, this work may add not so much to the
understanding of the process itself, was it not for a field-theoretic
re-formulation, which is particularly flexible and elegant. The price is
a process that ultimately differs from the original model. In hindsight,
the agreement of the original Wiener Sausage problem with the process
used here to formulate the problem field-theoretically deserves further scrutiny.
In the following, we first summarise our findings above with respect to the original Wiener Sausage problem, before discussing in further detail the field-theoretic insights.

\subsection{Summary of results in relation to the original Wiener
Sausage}\slabel{summary_results}
The original Wiener Sausage problem is concerned with the volume traced out by
a finite sphere attached to a Brownian particle. In the present analysis, this
has been replaced by a Brownian particle attempting to spawn immobile offspring
at Poissonian rate $\sigma$. The attempt fails if such immobile particles are
present already. 
On the lattice, this process amounts to a variant
of the number of distinct sites visited \cite{Torney:1986}. 

Above, the field-theoretic treatment has been carried out perturbatively to one
loop for dimensions $d<2=d_c$, but it turns out that there are no higher order
loops to be considered. In any dimension, by construction and as a matter of universality, the large time and
space asymptotes of the original Wiener Sausage,
the process on the lattice and the field theory are expected to
coincide at least as far as exponents
are concerned. 

The tree level of the field theory describes the phenomenon without
interaction, \ie ignoring returns. The resulting observables are the asymptotes
of the Wiener Sausage volume in dimensions above $d=2$.  The moments found in
the bulk, Eqs.~\eref{V1_bulk}, \eref{V2_bulk}, \eref{V3_bulk},
\eref{higher_bulk_moments} and generally \eref{general_scaling_tree}, $\ave{V^m} \propto m! \tau\sigma^{m-1} r^{-m}$, 
coincide with those from the exact moment
generating function \Eref{MGF_tree_level} of the process ignoring return,
obtained by probabilistic considerations. 

In the infinite slab, the field
theory still produces exact results (of the process ignoring return), such as
Eqs.~\eref{V_open_details} and \eref{V2_tree_open}, although higher moments are
tedious to calculate in closed form, \Eref{V3_tree_open}. Again, they are easily
verified using generating functions, such as \Eref{exact_residence_moments}, which 
also confirms the general form \Eref{general_scaling_open_tree},
$\ave{V^m}\propto \tau \sigma^{m-1} L^{2m} D^{-m}$, determined
field-theoretically.

Below two dimensions, infrared divergences occur in the perturbation theory,
which need to be controlled by a finite extinction rate $r$ (or $\epsilon'$).
It turns out that all orders can be dealt with at once, because ``parquet diagrams''
\cite{Herbut:2007} can be summed over in a geometric (Dyson) sum, such as
\Eref{kappa_renormalises_itself}. We can therefore expect exact universal exponents
of asymptotes, whereas amplitudes are generally non-universal and can be
affected by field-theoretically irrelevant terms.  In the bulk, the 
asymptotes Eqs.~\eref{V_bulk_r}, \eref{V_bulk_t},
\eref{V2_bulk_r} and generally \eref{general_scaling}, 
$\ave{V^m} \propto m! \tau \sigma^{m-1} r^{-md/2} (D^{d/2}/\kappa)^{m}$,
reproduce the (leading order) exponents as known in the literature
\cite{BerezhkovskiiMakhnovskiiSuris:1989}. In one dimension, the first moment
of the volume, \Eref{V_t_exact}, reproduces the asymptote (in large $t$) in the
continuum \cite{BerezhkovskiiMakhnovskiiSuris:1989} and on the lattice
\cite{Montroll:1964,Torney:1986}. Even the amplitude is reproduced correctly.

The bulk calculations can be modified to apply to the infinite slab, producing
\Eref{general_scaling_open}, $\ave{V^m} \propto m! \tau \sigma^{m-1}
(L/\pi)^{md} \kappa^{-m}$.  However, the renormalisation in
this case is correct only to leading order in $\epsilon$, as terms of order $\epsilon^0$, such as
\Eref{polylogarithm}, were omitted (whereas in the bulk, the 
$Z$-factor was exact, \Eref{Z_exact_d_1} or \Eref{Z_exact}). In one dimension, \ie
when the walker can explore only a finite interval, the amplitude of the first
moment for uniformly distributed initial starting points, \Eref{V_open_L} at $d=1$, coincides with the exact result,
\Eref{V_open_exact}. However, placing the particle
initially at the centre results in an amplitude, \eref{V_open_centre}, that
differs from the exact result.

Unless one is prepared to allow for a space-dependent $\kappa$ (whose
space dependence is in fact irrelevant in the field-theoretic sense) as suggested in \Eref{one_loop_open_next_step} for the infinite slab,
one cannot expect the resulting amplitudes to recover the exact
results. That \Eref{V_open_exact} does so nevertheless, may be explained by
the ``averaging effect'' of the uniform driving, given that 
\[
\int_0^L \dint{z}
\left(
\Li_{1+\epsilon}\left(\exp{\frac{2\pi \imag z}{L}}\right)
+
\Li_{1+\epsilon}\left(\exp{-\frac{2\pi \imag z}{L}}\right)
\right)
= 
0 \ ,
\]
see \eref{open_sum}.

As alluded to above, the field-theoretic description of the Wiener
Sausage is very elegant, not least because the diagrams have an
immediate interpretation. For example, 
$\tikz[baseline=-2.5pt]{
\draw[tactivity](0.4,0) -- (0,0);
\draw[tactivity](0,0) -- (-0.4,0);
\draw[tsubstrate] (-140:0.4) -- (0,0);
}$ 
corresponds to a substrate particle deposited while the active particle
is propagating. Correspondingly, 
$\tikz[baseline=-2.5pt]{
\draw[tactivity] (0,0) -- (-0.3,0);
\draw[tsubstrate] (-140:0.4) -- (0,0);
\draw[tactivity] (1.3,0) -- (1,0);
\draw[tactivity] (1,0) -- (0,0);
\draw[tsubstrate] (0,0) to[bend right,in=-140,out=-40] (1.,0);
}$ is the suppression of a deposition as the active particle encounters
an earlier deposition --- the active particle
\emph{returns} to a place it has been before. All loops can therefore be
contracted along the wavy line, 
$\tikz[baseline=-2.5pt]{
\draw[tsubstrate] (-0.4,0) -- (0,0);}$,
to produce a \emph{trajectory}, say
$\tikz[baseline=-2.5pt]{
\draw[tAactivity] (0,0) -- (-0.3,0);
\draw[tAactivity,rounded corners] (0,0) to[bend right,in=-140,out=-40] (0,0.5) to[bend
right,in=-140,out=-40] (0,0);
\draw[tAactivity] (0.3,0) -- (0,0);
\draw[tsubstrate] (-140:0.4) -- (0,0);
}$ or more strikingly just
$\tikz[baseline=-2.5pt]{
\draw[tAactivity] (0,0) -- (-0.3,0);
\draw[tAactivity,rounded corners] (0,0) to[bend right,in=-140,out=-40] (0,0.5) to[bend
right,in=-140,out=-40] (0,0);
\draw[tAactivity] (0.3,0) -- (0,0);
}$, illustrating that the loop integrals calculated above, in fact capture
the probability of a walker to return: $W\propto\omega^{-\epsilon/2}$,
\Eref{def_W}, which in the time domain gives 
$t^{-d/2}$.

\subsection{Original motivation}
The present study was motivated by a number of ``technicalities'' which
were encountered by one of us during the study of a more complicated
field theory. The \textbf{first issue}, as mentioned in the introduction, was the
``fermionic'' or excluded-volume interaction. In a first step, that was
generalised to an arbitrary carrying capacity $n_0$, whereby the deposition
rate of immobile offspring varies smoothly in the occupation number
until the carrying capacity is reached. 
It was argued above,
\Fref{funny_lattice}, that the constraint to a finite but large carrying
capacity $n_0$, which may be conceived as less brutal than setting 
$n_0=1$, can be understood as precisely the latter constraint, but on a
more complicated lattice. 

Even though the field theory was constructed
in a straight-forward fashion, the perturbative implementation of the constraint,
namely by effectively discounting depositions that should not have
happened in the first place, make it look like a minor miracle that it
produces the correct scaling (and even the correct amplitudes in some
cases). We conclude that the present
approach is perfectly suitable to implement excluded volume constraints.

It is interesting to vary $n_0$ in the expressions obtained for the
volume moments. At first it may not be obvious that, for example, the
first volume moments in one dimension, \Eref{V_t_exact} and \Eref{V_open_exact}, are linear in
$n_0$, because $\kappa=\tau/n_0$, \Eref{bare_level_couplings}. Given
that $\kappa$ enters the $m$th moment $\ave{V^m}$ as $\kappa^{-m}$, 
\Eref{general_scaling} and \Eref{general_scaling_open}, the carrying capacity
therefore enters through $\kappa=\gamma/n_0$ as $n_0^{m}$. Even though
the carrying capacity enters smoothly into the deposition rate (or,
equivalently, the suppression of the deposition), in dimensions $d<2$ each site is visited
infinitely often and is therefore ``filled up to the maximum'' with
offspring particles, as if the carrying capacity was a hard cutoff (\ie as if the deposition rate were constant until the occupation reaches the
carrying capacity). The volume of each sausage therefore increases by a
factor $n_0$ in dimensions $d<2$ and is independent of it (as $\kappa$
does not enter) in $d>2$.

The \textbf{second issue} to be investigated was the presence of open boundaries.
This is, obviously, not a new problem as far as field theory is
concerned in general, but in the present case being able to change
boundary conditions exploits the flexibility of the field-theoretical
re-formulation of the Wiener Sausage and allows us to probe results in
a very instructive way. 

It is often said that translational invariance corresponds to momentum
conservation in $\kvec$-space, but the present study highlights some subtleties.
As far as bare propagators are concerned, open,
periodic, or, in fact, reflecting boundary conditions all allow it to be
written with a Kronecker-$\delta$ function. In that sense, \emph{bare} propagators
do not lose momentum. Momentum, however, is generally not conserved in
vertices, \ie vertices with more than two legs do not come with a simple
$\delta_{n+m+\ell,0}$, but rather in a form such as \Eref{def_epsilon_nml} or \Eref{def_U}.

These more complicated expressions are present even at tree level,
\Eref{V2_tree_open}. This touches on an interesting feature,
namely that non-linearities are present even in
dimensions above the upper critical dimension --- they have to, as
otherwise the tree level lacks a mechanism by which immobile offspring
are deposited.

Below the upper critical dimension, the lack of momentum conservation
has three major consequences: Firstly, each vertex comes with a summation
and so a loop formed of two vertices, \Eref{one_loop_open}, requires not only one
summation ``around the loop'' but a second one accounting for another
index, which is no longer fixed by momentum conservation. This is a
technicality, but one that requires more and potentially serious computation. Secondly,
and more seriously, the very structure of the vertex might change. For
example, at bare level $\kappa$ comes with a factor $LU_{n,m+k,\ell}$,
but that $U_{n,m+k,\ell}$ might change under renormalisation. 

Finally, the third and probably most challenging consequence is the loss
of momentum conservation in the propagator. While a lack of
translational invariance may not be a problem at bare level, the
presence of non-momentum conserving vertices can render the propagators
themselves non-momentum conserving --- provided the propagators
renormalise at all (see the discussion after \Eref{full_tau_open}), which they do not in the present case, as far as the
two shown in \Eref{bare_activity_propagator} are concerned. However, 
$\tikz[baseline=-2.5pt]{
\draw[tactivity] (0.4,0) -- (0,0);
\draw[tsubstrate] (0,0) -- (-0.4,0);
}$
parameterised by $\tau$
has every right to be called a propagator and it does renormalise.
Luckily, however, it never features within loops, so the complications
arising from its new structure can be handled within observables and does
not spoil the renormalisation process itself.

A consequence of the Dirichlet boundary conditions is the existence of a
lowest, non-vanishing mode, $q_1=\pi/L$, \Eref{general_scaling_open}, which, in fact, turns out to
play the r{\^o}le of the effective mass --- just like the minimum of the 
inverse propagator, $(-\imag\omega+D\kvec^2+r)$, the ``gap'', is $r$ in
the bulk, it is $Dq_1^2+r$ in the presence of Dirichlet boundary
conditions, and thus does not vanish even when $r=0$. This is a nice
narrative, which is challenged, however, when periodic boundary
conditions are applied. At tree
level, when the interaction is switched off, periodic boundaries cannot
be distinguished from an infinite system, and so we would evaluate
at tree level an infinite and a periodic system both at $\kvec=0$ and
$k_n=0$ respectively, producing exactly the same expectation (for exactly the right reason).

The situation is different beyond tree level. Periodic or open, the
system is finite. However,
periodic boundaries do not drain active particles, so the lowest wave
number vanishes, $k_n=0$. To control the infrared (in the infinite directions), a finite extinction rate $r$
is necessary, which effectively competes with the system size $L$ via
$\rhotilde\propto L^2 r/D$, \Eref{one_loop_cyl_final_step} and
\Eref{one_loop_cyl_final_step_large_circ}. If $\rhotilde$ is large, bulk
behaviour $\propto \rhotilde^{-\epsilon/2}$ is recovered,
\Eref{one_loop_cyl_final_step_large_circ}, as is the case in the open system (see footnote \fnref{open_recovers_bulk} before \Eref{open_sum}). For
moderately small values, $\zeta(3-d)\propto 1/\epsilon$ dominates,
\Eref{cyl_sum_small}, a signature of a $d$-dimensional system with \emph{open}
boundaries, \Eref{one_loop_open_final_step}. In that case, scaling amplitudes are in fact $\propto
L^{\epsilon}$, \Eref{one_loop_cyl_final_step_simplified}. However, the
presence of the $0$-mode allows for a different asymptote as $\rhotilde$
is lowered further, the bulk-like term governing the $d-1=\dtilde$
infinite dimensions takes over, $\propto L^{-1}((r+\epsilon'-\imag\omega)/D)^{-\epsilontilde/2}$.
It is the appearance of that term and \emph{only} that term
which distinguishes periodic from open boundary conditions. 

So, the narrative of ``lowest wave number corresponds to mass'' is
essentially correct. In open systems, it dominates for all small masses.
In periodic systems, the scaling of the lowest non-zero mode competes with that of a $d-1$-dimensional bulk system due to the presence of a $0$-mode in the periodic dimension, which asymptotically drops out.

The \textbf{third point} that was to be addressed in the present work
were the special properties of a propagator of an immobile species. The
fact that the propagator is, apart from $\delta(\kvec+\kvec')$,
\Eref{substrate_propagator},
independent of the momentum is physically relevant as the particles
deposited stay where they have been deposited and so the walker has to truly
return to a previous spot in order to interact. Also, deposited particles are not
themselves subject to any boundary conditions --- this is the reason for
the ambiguity of the eigenfunctions that can be used for the fields of
the substrate particles. If deposited particles were to ``fall off'' the lattice, 
the volume of the sausage on a finite lattice cannot be determined by
taking the $\omega\to0$ limit.

It is interesting to see what happens to the crucial integral
\Eref{one_loop}
when the immobile propagator is changed to $(-\imag\omega+\nu\kvec^2+\epsilon')^{-1}$:
\begin{multline}\elabel{one_loop_with_extras}
\kappa^2 \int \ddintbar{k'}\dintbar{\omega'}
\frac{1}{-\imag\omega'+\nu\kvec^2+\epsilon'}
\frac{1}{-\imag(\omega-\omega')+D(\kvec-\kvec')^2+r}\\
=
\frac{\kappa^2}{(4\pi)^{d/2}(D+\nu)}\left(\frac{r+\epsilon'-\imag\omega+\frac{D\nu}{D+\nu}\kvec^2}{D+\nu}\right)^{-\epsilon/2}
\Gamma\left(\frac{\epsilon}{2}\right) \ ,
\end{multline}
which at external momentum $\kvec=0$ is \Eref{one_loop} with $D$
replaced by $D+\nu$.
The integral thus remains essentially unchanged, just that the effective diffusion
constant is adjusted by $D\to D+\nu$.

A slightly bigger surprise is the fact that $\epsilon'$, the IR
regulator of the substrate propagator, is just as good an IR regulator
as $r$, the IR regulator of the activity propagator. The entire field
theory and thus all the physics discussed above, does not change when
the ``evaporation of walkers'' is replaced by ``evaporation of substrate
particles''. The stationarity of both in infinite systems is obviously
due to two completely different processes, which, however, have the same
effect on the moments of the Sausage Volume: If $r$ is finite, then
a walker eventually disappears, living behind the trace of substrate
particles, which stay indefinitely. If $\epsilon'$ is finite, then
stationarity is maintained as substrate particles disappear while new
ones are produced by an ever wandering walker.

Finally, the \textbf{fourth issue} to be highlighted in the present work
was that of observables which are spatial integrals of densities. These
observables have a number of interesting features. As far as space is
concerned, eigenfunctions with a $0$-mode immediately give access to
integrals over all space. However, open boundaries force us to perform a
summation (and an awkward looking one, too, say \Eref{V_open_details}).

\subsection{Future work}
Two interesting extensions of the present work deserve brief 
highlighting. Firstly, the Wiener Sausage may be studied on networks:
Given a network or an ensemble thereof, how many distinct sites are
visited as a function of time. The key ingredient in the analysis is the
lattice Laplacian, which provides a mathematical tool to describe the
diffusive motion of the walker. The contributions $\kvec^2$ and $q_n^2$
in the denominator of the propagator, \Eref{bare_activity_propagator}
and \Eref{bare_activity_propagator_open}, 
are the squared eigenvalues of the Laplacian operator in the continuum
and, in fact, of the lattice Laplacian, for, say, a square
lattice. The integrals in $\kvec$-space and, equivalently, sums like
\Eref{convenient_choices} and \Eref{V_open_details} should be seen as
integrating over all eigenvalues $\kvec^2$, whose density in $d$
dimensions is proportional 
to $|\kvec|^{d-1}$. It is \emph{that} $d$ which determines the scaling
in, say, $\ave{V} \propto t^{d/2}$ for $d<2$. In other words, if
$|\kvec|^{d_s-1}$
is the density of eigenvalues (the density of states) of the lattice Laplacian, then the Wiener
Sausage volume scales like $t^{d_s/2}$ (and the probability of return
like $t^{-d_s/2}$). 
Provided the propagator does not acquire an anomalous dimension, which could depend on $d_s$ in
a complicated way, the difference between a field theory on a regular
lattice with dimension $d$ and one on a complicated graph with spectral dimension
$d_s$ is captured by replacing $d$ by $d_s$ \cite[p.~23]{Dorogovtsev:2010}.
We confirmed this finite size scaling of the Wiener Sausage on four different fractal lattices.

The second interesting extension is the addition of processes, such as
branching of the walkers itself. In that case they not only interact
with their past trace, but also with the trace of ancestors and successors.
This field theory is primarily dominated by the branching ratio, 
say $s$,
and 
$\lambda$,
whereas $\kappa$, $\chi$ and $\xi$ are irrelevant.
Preliminary results suggest that $d_c=4$ \cite[see
also]{vandenBergBolthausendenHollander:2004} in this case and again
$\ave{V}\propto L^{2-\epsilon}$, this time, however, with
$\epsilon=4-d$. Higher moments seem to follow 
$\ave{V^m}\propto L^{(m-1)d+2-\epsilon}=L^{md-2}$. The latter result
suggests that the dimension of the cluster formed of sites visited is
that of the underlying lattice.

\begin{acknowledgement}
The authors would like to thank Andy Thomas and Niall Adams for computing support,
Hoai Nguyen Huynh for the numerical results on the Wiener Sausage on fractals and
Aman Pujara and
Songhongyang Yuan for computational support. 
SN would like to thank Imperial College and in particular GP for their hospitality and the German National Academic Foundation for its generous support of SN's stay at Imperial College. 
This work has been supported by the EPSRC Mathematics Platform grant EP/I019111/1.
\end{acknowledgement}

\end{document}